\documentclass[english]{article}%
\usepackage{amssymb}
\usepackage{fontenc}
\usepackage[latin1]{inputenc}
\usepackage{amsmath}
\usepackage{fontenc}
\usepackage{babel}
\usepackage{amsfonts}
\usepackage{graphicx}
\usepackage{geometry}%
\usepackage{cite}
\providecommand{\U}[1]{\protect\rule{.1in}{.1in}}
\begin{document}

\title{States of charged quantum fields and their statistical properties in the
presence of critical potential steps}
\author{S.P. Gavrilov$^{1,2}$%
\thanks{gavrilovsergeyp@yahoo.com;gavrilovsp@herzen.spb.ru}, D.M.
Gitman$^{2,3,4}$\thanks{gitman@if.usp.br}, and A.A. Shishmarev$^{4}%
$\thanks{a.a.shishmarev@mail.ru}\\
{\normalsize $^{1}$ Department of General and Experimental Physics, }\\
{\normalsize Herzen State Pedagogical University of Russia, Moyka Embankment 48, 191186, St. Petersburg, Russia;}\\
{\normalsize $^{2}$ Department of Physics, Tomsk State University, 634050 Tomsk, Russia;} \\
{\normalsize $^{3}$ P.N. Lebedev Physical Institute, 53 Leninsky Prospekt,
119991 Moscow, Russia;}\\
{\normalsize $^{4}$ Institute of Physics, University of S{\"a}o Paulo, CEP
05508-090, S{\"a}o Paulo, S{\"a}o Paulo, Brazil;}}
\maketitle

\begin{abstract}
The evolution of charged quantum fields under the action of constant nonuniform
electric fields is studied. To this end we construct a special generating
functional for density operators of the quantum fields with different initial
conditions. Then we study some reductions of the density operators, for
example, reductions to electron or positron subsystems, reduction induced by
measurements, and spatial reduction to the left or to the right subsystems of
final particles. We calculate von Neumann entropy for the corresponding
reduced density operators, estimating in such a way an information loss. Then
we illustrate the obtained results by calculations in a specific background of
a strong constant electric field between two infinite capacitor plates
separated by a finite distance $L$.

\emph{Keywords}: QED with $x$-electric potential steps, particle creation,
von Neumann entropy\newline
\quad~ PACS number(s): 12.20.Ds, 03.65.Ud

\end{abstract}

\maketitle

\section{Introduction}

Problems of quantum field theory with external backgrounds violating the
vacuum stability have been studied systematically for a long time.
Recently, they have drawn special attention due to new real
possible applications in astrophysics and physics of nanostructures. In
these areas one often encounters a situation where the effects of vacuum
instability (in particular, due to the presence of potential steps, that is,
inhomogeneous electric fields) and finite temperature are combined.
Astrophysical objects such as black holes and neutron stars can generate
huge electromagnetic fields in their vicinity. The Coulomb barrier at the
quark star surface of a hot strange star may be a powerful source of $%
e^{+}e^{-}$ pairs, which are created in extremely strong constant electric
fields (dozens of times higher that the critical field $E_{c}$) of the
barrier, and they flow away from the star (see \cite{Web+etal14} for the
review). Such emission may be a good observational signature of bare strange
stars. The existence of critical electric fields on the quark star surfaces
was also predicted in Ref.~\cite{Alf+etal01} in the transition at very
high densities, from the normal nuclear matter phase at the core to the
color-flavor-locked phase of quark matter at the inner core of hybrid stars.
The possibility of existence of critical electromagnetic fields at the core
surface of a neutron star was indicated in Ref.~\cite{Ruf+etal11}.
Critical electric fields are expected to appear in the late phases of
gravitational collapse and from cosmological horizons, with a consequent
process of pair creation by vacuum polarization (see, e.g., reviews in \cite%
{RufVSh10,CHM08,AndMot14}). There is a close connection between particle
creation by strong electrostatic potentials, in particular, by steps and
barriers, and the Unruh effect, which is the phenomenon of particle emission
from black holes and cosmological horizons. Particle creation from the
vacuum by external fields (the generation of electron-hole pairs by the
electric field or Zener tunneling) has become an observable effect in
physics of graphene and similar nanostructures (e.g., in topological
insulators and Weyl semimetals); this area is currently under intense
development (see the reviews in \cite{dassarma,VafVish14,GelTan15} as well as the recent
article \cite{Akal+etal19} and references therein).

Note that the cases with homogeneous time-dependent electric fields are
considered in most of these articles. The effect of pair production from the
vacuum by time-dependent electric fields was considered in a number of
works, starting with the pioneer work of Schwinger \cite{Schwinger}, followed by that of
Nikishov \cite{Nikishov1,Nikishov2}, Brezin and Itzykson \cite{Berezin}, and
many others. Later a nonperturbative formulation of QED with so-called $t$-electric potential steps 
(time-dependent potentials of special form) was
developed in Refs.~\cite{Gitman} and applied to various physical problems (
see, e.g., Refs.~\cite{143,density,DvGavGi,20}). In particular, quantum
entanglement in the Schwinger effect of Dirac or the Klein-Gordon field due to
the $t$-electric steps, between a subsystem and the rest of the system, as
measured by the von Neumann entropy of the reduced density matrix, was
calculated \cite{stat} (see as well Refs.~\cite{MizEb14,LiShi17}). For
more information on the subject see recent reviews in \cite%
{Dun09,RufVSh10,GelTan15,t-case}, where the progress on particle creation
due to time-dependent field configurations is described and a number of
important applications of such fields are considered.

However, the case where external backgrounds are represented by strong
time-independent nonuniform electric fields concentrated in restricted space
areas is much closer to a real experimental situation. We refer to such
backgrounds as $x$-electric potential steps. There are theoretical
articles where fields of this type are considered; see, for example, Refs.~%
\cite{Nikishov2,Nikishov3,Greiner,Sauter,Kim1,Kim2,Gies,Chern1,Chern2,Wang}.
In the recent work \cite{x-case} a consistent nonperturbative (with respect
to an external electric field in zeroth order in the radiation interaction)
formulation of QED with $x$-electric potential steps strong enough to
violate the vacuum stability was constructed. In Refs.~%
\cite{unitarity,x-case} some quantum effects related to a violation of the vacuum
instability by $x$-electric potential steps were calculated. The particle
creation effect is crucial for understanding the conductivity of graphene,
especially in the so-called nonlinear regime. In this regime it is natural
to consider a constant voltage applied between two electrodes. Possible
experimental configurations for testing the pair creation by a linear step
of finite length were proposed in Ref.~\cite{allor}. For the case of a
constant voltage between two electrodes the evidence of the existence of
electron-hole pair creation was obtained in graphene by its indirect
influence on the graphene conductivity \cite{Van+etal10}. The first
experimental observation of graphene optical emission induced by the intense
terahertz pulse was recently reported \cite{Olad+etal17}. The
experimental data are in a good agreement with the theory of Landau-Zener
interband transitions. The impact of Zener tunneling on the charge-transport
properties of graphene in the high-field regime was studied theoretically in
Ref. \cite{KaneLM15}. It is shown that the inclusion of both Zener tunneling
and electron-electron relaxation improves the agreement with the
measurements performed in graphene in the high-field regime at low doping.
The {\textit{p-n}} junctions and sharp {\textit{n-n}} junctions can also play the role of
potential steps. For these steps the Klein tunneling was observed by several
experimental groups (but only for the kinetic energies of electron that
exclude the possibility of pair production) (see, e.g., the review in \cite%
{dassarma}). It should also be noted that in the context of strong
interactions and quantum chromodynamics, a similar phenomenon may play a role
in the discussion of particle production in heavy-ion collisions or in the
decay of \textquotedblleft hadronic strings\textquotedblright\ in the
process of hadronization (see Ref.~\cite{GelTan15} for a review).

In this article we study the evolution of different initial states of charged
quantum fields in $x$-electric critical potential steps, using the above-mentioned formulation of QED \cite{x-case}. 
To this end, we construct
density operators for different initial states of the system of quantum
fields. We consider pure initial states and thermal (mixed) initial states.
Corresponding final states are studied using three types of reductions. Since 
$x$-electric potential steps cause a natural division of created particles in
subsystems of electrons and positrons substantially separated spatially, we
first consider reductions to electron or positron subsystems. In the
background under consideration, it is interesting to calculate reductions to
the left and right parts of the whole system and compare the obtained states
with states resulting from the previously mentioned reductions. Finally, we
study reductions due to possible measurements of a number of final particles.
The latter kind of reductions can also occur due to some decoherence
processes, such as collisions with some external sources (e.g. with
impurities in the graphene). To study the loss of the information in all the
reductions, we calculate von Neumann entropy for reduced density operators.
In two first reduction cases this entropy can also be identified with a
measure of quantum entanglement between the corresponding quantum
subsystems. The article is organized as follows. In Sec. \ref{general} we
recall basic points of QED with $x$-electric potential steps. In Sec. \ref%
{initial} we present density operators for different initial states of
charged quantum fields. All the above mentioned reductions are presented in
Sec. \ref{reduced}. The corresponding von Neumann entropy is calculated
in Sec. \ref{Neumann}. Special generating functionals allow us to
construct density matrices for different initial conditions by choosing
appropriate sources presented in Sec. \ref{initial}, and their normal
forms are placed in Appendix A. In Appendix B we briefly consider
the case when the initial state of the system is given by a pure state with
a definite number of particles. Some useful operatorial relations are given
in Appendix C.

\section{QED with $x$-electric potential steps\label{general}}

The general theory of quantization of charged fields in the presence of
critical potential steps that we use was formulated by Gitman and Gavrilov
in Ref. \cite{x-case}. They constructed a special self-consistent QED
with $x$-electric potential steps utilizing the so-called generalized Furry
picture. In the framework of this QED it is possible to take into account
the external electric field exactly in zeroth order in the
radiation interaction when the analytical solutions of the Dirac equation in
the corresponding field are known. Here we repeat some crucial moments of
this theory. In this article we generally adapt the notation used in Ref. \cite%
{x-case}; we utilize the system of units where $c=\hbar =1$.

\begin{figure}[th]
\begin{center}
\includegraphics[width=0.6\textwidth]{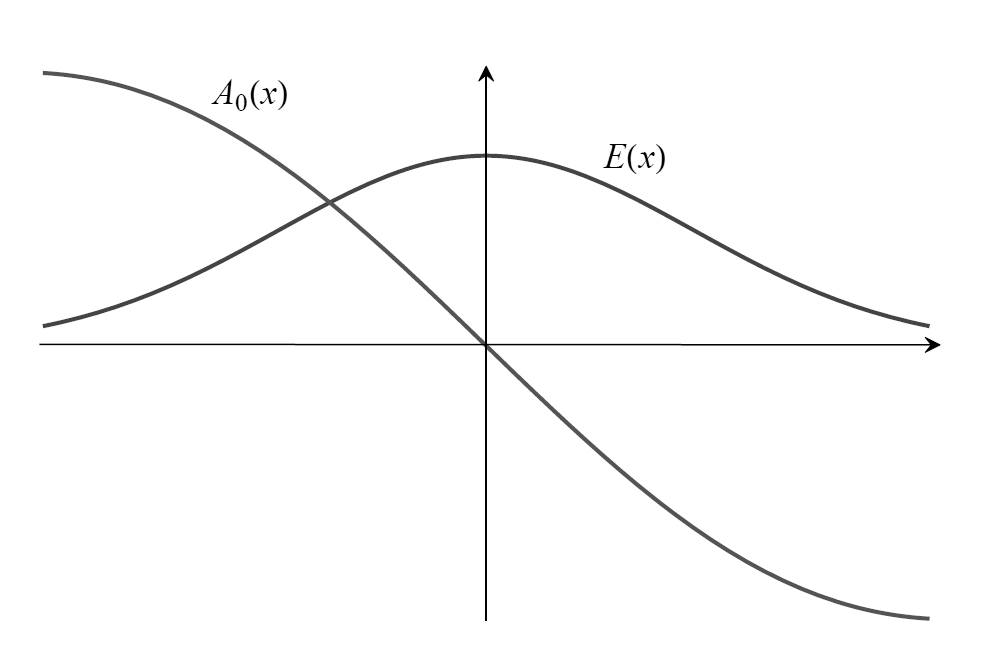}
\caption{Example of the $x$-potential electric step.}
\label{fig.1}
\end{center}
\end{figure}

We work in ($d=D+1$)-dimensional Minkowski space-time parametrized by
coordinates $X$,%
\begin{equation}
X=\left( X^{\mu },\ \mu =0,1,\ldots ,D\right) =\left( t,x,\mathbf{r}_{\bot
}\right) ,\ \ X^{0}=t,\ x=X^{1},\ \mathbf{r}_{\bot }=\left( X^{2},\ldots
,X^{D}\right) ,  \label{gen.00}
\end{equation}%
which correspond to an $x$-electric potential step of the form
\begin{equation}
A^{\mu }(X)=\left( A^{0}(x),A^{j}=0,\ j=1,2,\ldots ,D\right) ,
\label{gen.00a}
\end{equation}%
so that the magnetic field $B$ is zero and the electric field $E$ reads 
\begin{equation}
\mathbf{E}(X)=\mathbf{E}(x)=\left( E_{x}(x),0,\ldots ,0\right) ,\ 
E_{x}(x)=-A_{0}^{\prime }(x)=E(x).  \label{gen.00b}
\end{equation}%
The electric field (\ref{gen.00b}) is directed along the $x$ axis, it is
inhomogeneous in the $x$ direction, and does not depend on time $t$. The
main property of any $x$-electric potential step is 
\begin{equation}
A_{0}(x)\overset{x\rightarrow \pm \infty }{\rightarrow }A_{0}(\pm \infty ),\ \
E(x)\overset{\left\vert x\right\vert \rightarrow \infty }{\rightarrow }0,
\label{gen.00c}
\end{equation}%
where $A_{0}(\pm \infty )$ are constant quantities, which means that
the electric field under consideration is switched off at spatial infinity.
In addition, it is supposed that the first derivative of the scalar
potential $A_{0}(x)$ does not change its sign for any $x\in R$, and that there exist points $x_{\mathrm{L}}$\ and $x_{\mathrm{R}}$\
($x_{\mathrm{R}}>x_{\mathrm{L}}$) such that for $x\in S_{\mathrm{L}%
}=(-\infty ,x_{\mathrm{L}}]$\ and for $x\in S_{\mathrm{R}}=[x_{\mathrm{R}%
},\infty )$\ the electric field is already switched off, so%
\begin{eqnarray}
\left. A_{0}(x)\right\vert _{x\in S_{\mathrm{L}}} &=&A_{0}(-\infty ),\ \
\left. E(x)\right\vert _{x\in S_{\mathrm{L}}}=0,  \notag \\
\left. A_{0}(x)\right\vert _{x\in S_{\mathrm{R}}} &=&A_{0}(+\infty ),\ \
\left. E(x)\right\vert _{x\in S_{\mathrm{R}}}=0,  \label{add.1}
\end{eqnarray}%
whereas the electric field is not zero in the region $S_{\mathrm{int}}=(x_{%
\mathrm{L}},x_{\mathrm{R}})$\ (note that both $x_{\mathrm{L}}$\ and $x_{%
\mathrm{R}}$\ can tend to infinity). An example of an $x$-potential step can
be found in Fig. \ref{fig.1}. 

There are two types of electric steps, noncritical and critical, which are
distinguished by their magnitudes 
\begin{equation}
\mathbb{U}=\left\{ 
\begin{array}{l}
\mathbb{U}<2m,\ \ \mathrm{noncritical\ steps} \\ 
\mathbb{U}>2m,\ \ \mathrm{critical\ steps}%
\end{array}%
\right. ,\ \   \label{add.2}
\end{equation}%
where $\mathbb{U}=U_{\mathrm{R}}-U_{\mathrm{L}}$, $U_{\mathrm{L (R)}}$ is the left (right) asymptotic potential energy, $U_{\mathrm{L}%
}=U(x\rightarrow -\infty )$, $U_{\mathrm{R}}=U(x\rightarrow +\infty )$, $U(x)=-eA_{0}(x)$ is the potential energy of the electron in the $x$-electric
potential step, and $m$ is the electron mass. We are mostly interested in
critical steps, which can produce pairs from the vacuum.

\begin{figure}[th]
\includegraphics[width=\textwidth]{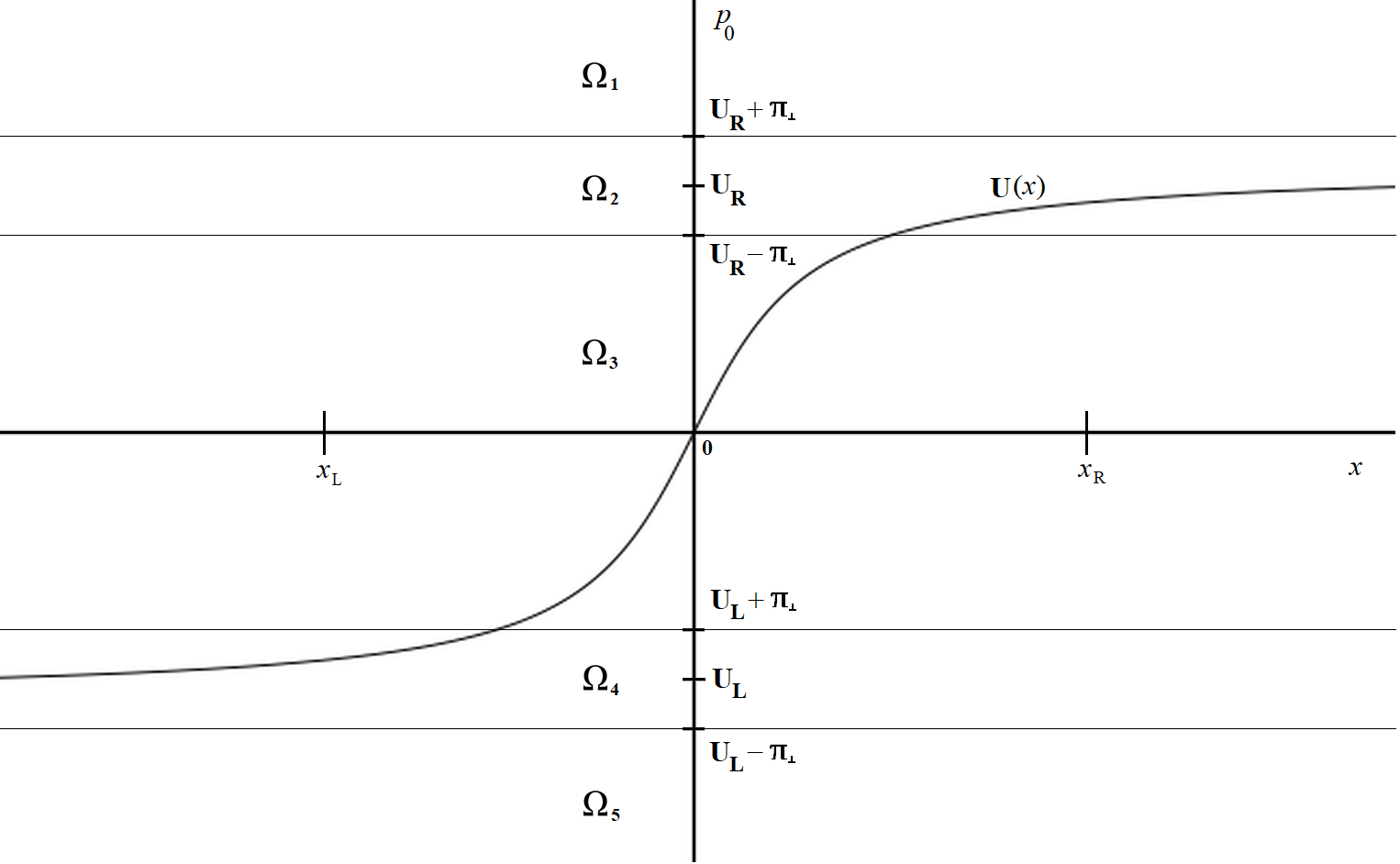}
\caption{Graphical representation of the quantum ranges $\Omega_i$ showing the potential energy of the electron $U(x)=-eA_{0}(x)$ in the
electric field.}
\label{fig.2}
\end{figure}

One of the most important points of QED with critical potential steps is
that the whole space of quantum numbers $n\in \Omega $ [which are the full
energy of particle $p_{0}$, momenta $\mathbf{p}$, and spin $\sigma $, $%
n=(p_{0},\mathbf{p},\sigma )$] can be divided into five different ranges $%
\Omega _{i}$, $i=1,\ldots ,5$, $n_{i}\in \Omega _{i}$, where the solutions
of the corresponding Dirac equation have similar forms. The full operator of the
Dirac field can be presented as a sum of operators defined for each
particular range $\Omega _{i}$,%
\begin{equation}
\hat{\Psi}(X)=\sum_{i=1}^{5}\hat{\Psi}_{i}(X).  \label{gen.0}
\end{equation}%
The explicit forms of operators $\hat{\Psi}_{i}(X)$ are given in Appendix A,
and a graphical representation of the quantum ranges $\Omega _{i}$ can be
found in Fig. \ref{fig.2}. 
Detailed consideration of each range $\Omega _{i}$ was carried out in
Ref.~\cite{x-case}; here we repeat only the most important points. In these
ranges there exist two types of solutions of the Dirac equation, $_{\zeta
}\psi _{n}(X)$ and $^{\zeta }\psi _{n}(X)$, $\zeta =\pm $. Those solutions
satisfy the following asymptotic conditions:%
\begin{eqnarray}
&&_{\zeta }\psi _{n}(X)=\exp \left( -ip_{0}t+i\mathbf{p}_{\bot }\mathbf{r}%
_{\bot }\right) \ _{\zeta }\varphi _{n}(x),\ \ _{\zeta }\varphi _{n}(x)=\
_{\zeta }\varphi _{n}^{\mathrm{L}}(x),\ \ x\in S_{\mathrm{L}},  \notag \\
&&\left[ \hat{p}_{x}^{2}-\left( p_{0}-U_{\mathrm{L}}\right) ^{2}+\pi _{\bot
}^{2}\right] \ _{\zeta }\varphi _{n}^{\mathrm{L}}(x)=0,\ \ \pi _{\bot
}^{2}=m^{2}+\mathbf{p}_{\bot }^{2},  \notag \\
&&^{\zeta }\psi _{n}(X)=\exp \left( -ip_{0}t+i\mathbf{p}_{\bot }\mathbf{r}%
_{\bot }\right) \ ^{\zeta }\varphi _{n}(x),\ \ ^{\zeta }\varphi _{n}(x)=\
^{\zeta }\varphi _{n}^{\mathrm{R}}(x),\ \ x\in S_{\mathrm{R}},  \notag \\
&&\left[ \hat{p}_{x}^{2}-\left( p_{0}-U_{\mathrm{R}}\right) ^{2}+\pi _{\bot
}^{2}\right] \ ^{\zeta }\varphi _{n}^{\mathrm{R}}(x)=0,\ \ \hat{p}%
_{x}=-i\partial _{x}.  \label{add.3}
\end{eqnarray}%
Nontrivial solutions $^{\zeta }\psi _{n}(X)$ and $_{\zeta }\psi _{n}(X)$
exist only for quantum numbers $n$ that obey the relations%
\begin{eqnarray}
&&\left( p_{0}-U_{\mathrm{R}}\right) ^{2}>\ \pi _{\bot }^{2}\ \ \mathrm{for}%
\ \ ^{\zeta }\psi _{n}(X),  \label{add.3a1} \\
&&\left( p_{0}-U_{\mathrm{L}}\right) ^{2}>\ \pi _{\bot }^{2}\ \mathrm{\ for}%
\ \mathrm{\ }_{\zeta }\psi _{n}(X),  \label{add.3a2}
\end{eqnarray}%
and correspond to states with definite momenta $p^{\mathrm{R}}$\ and $p^{%
\mathrm{L}}$,%
\begin{eqnarray}
&&\hat{p}_{x}\mathrm{\ }_{\zeta }\psi _{n}(X)=p^{\mathrm{L}}\mathrm{\ }%
_{\zeta }\psi _{n}(X),\ \ x\rightarrow -\infty ,\ \ p^{\mathrm{L}}=\zeta 
\sqrt{(p_{0}-U_{\mathrm{L}})^{2}-\pi _{\bot }^{2}},  \notag \\
&&\hat{p}_{x}\ ^{\zeta }\psi _{n}(X)=p^{\mathrm{R}}\ ^{\zeta }\psi _{n}(X),\
\ x\rightarrow +\infty ,\ \ p^{\mathrm{R}}=\zeta \sqrt{(p_{0}-U_{\mathrm{R}%
})^{2}-\pi _{\bot }^{2}}.  \label{add.3b}
\end{eqnarray}

The ranges $\Omega _{1}$ and $\Omega _{5}$ exist for any step, critical or
noncritical, and are defined by the inequalities%
\begin{eqnarray}
&&p_{0}\geq U_{R}+\pi _{\bot }\ \ \mathrm{if\ }\ n\in \Omega _{1},  \notag \\
&&p_{0}\leq U_{L}-\pi _{\bot }\ \ \mathrm{if\ }\ n\in \Omega _{5},
\label{add.4}
\end{eqnarray}%
for a given $\pi _{\bot }$. Solutions $^{\zeta }\psi _{n}(X)$ can be
interpreted as either a wave function of an electron for $n\in \Omega _{1}$ or
a wave function of a positron for $n\in \Omega _{5}$ with the momenta $p^{%
\mathrm{R}}$ along the $x$ axis, whereas solutions $_{\zeta }\psi _{n}(X)$
can be interpreted as either a wave function of an electron for $n\in \Omega
_{1}$\ or a wave function of a positron for $n\in \Omega _{5}$ with momenta $%
p^{\mathrm{L}}$ along the $x$ axis.

The ranges $\Omega _{2}$ and $\Omega _{4}$ also exist for any step, and
include the quantum numbers $n\in \Omega _{2}$ that obey the inequalities 
\begin{eqnarray}
&&U_{\mathrm{R}}-\pi _{\bot }<p_{0}<U_{\mathrm{R}}+\pi _{\bot },\ (p_{0}-U_{%
\mathrm{L}})>\pi _{\bot }\text{\textrm{,}}\mathrm{\ \ if\ \ }2\pi _{\bot }<%
\mathbb{U},  \notag \\
&&U_{\mathrm{L}}+\pi _{\bot }<p_{0}<U_{\mathrm{R}}+\pi _{\bot },\mathrm{\ \
if\ \ }2\pi _{\bot }<\mathbb{U},  \label{add.5}
\end{eqnarray}%
and the quantum numbers $n\in \Omega _{4}$ that obey the inequalities%
\begin{eqnarray}
&&U_{\mathrm{L}}-\pi <p_{0}<U_{\mathrm{L}}+\pi ,\ \ (p_{0}-U_{\mathrm{R}%
})<-\pi _{\bot }\text{\textrm{,}}\mathrm{\ \ if\ \ }2\pi _{\bot }<\mathbb{U},
\notag \\
&&U_{\mathrm{L}}-\pi <p_{0}<U_{\mathrm{R}}-\pi ,\mathrm{\ \ if\ \ }2\pi
_{\bot }<\mathbb{U}.  \label{add.6}
\end{eqnarray}%
As a consequence of these inequalities there exist solutions $_{\zeta }\psi
_{n}(X)$, $n\in \Omega _{2}$, with definite left asymptotics and $^{\zeta
}\psi _{n}(X)$, $n\in \Omega _{4}$, with definite right asymptotics.
Solutions $_{\zeta }\psi _{n}(X)$, $n\in \Omega _{2}$, and $^{\zeta }\psi
_{n}(X)$, $n\in \Omega _{4}$, can be interpreted as wave functions of the
electron and positron, respectively. Nontrivial solutions $^{\zeta }\psi
_{n}(X)$, $n\in \Omega _{2}$, and $_{\zeta }\psi _{n}(X)$, $n\in \Omega _{4}$,
do not exist, as the inequality (\ref{add.5}) contradicts Eq. (\ref{add.3a1}) and the inequality (\ref{add.6}) contradicts Eq. (\ref{add.3a2}).

The range $\Omega _{3}$, the Klein zone, exists only for critical steps. The
quantum numbers $p_{\bot }$ are restricted by the inequality $2\pi _{\bot }<U$,
and for any of such $\pi _{\bot }$ quantum numbers $p_{0}$ obey the double
inequality%
\begin{equation}
U_{\mathrm{L}}+\pi \leq p_{0}\leq U_{\mathrm{R}}-\pi .  \label{add.7}
\end{equation}%
In the range $\Omega _{3}$\ there exist the following sets of solutions%
\begin{equation}
\left\{ \ _{\zeta }\psi _{n}(X)\right\} ,\ \ \left\{ \ ^{\zeta }\psi
_{n}(X)\right\} ,\ \ n\in \Omega _{3},\ \ \zeta =\pm .  \label{add.8}
\end{equation}%
However, the one-particle interpretation of these solutions based on the energy
spectrum in a similar way as has been done in the ranges $\Omega _{1}$\ and $%
\Omega _{5}$\ becomes inconsistent. Indeed, it is enough to see the
following contradiction: From the point of view of the left asymptotic area 
$S_{\mathrm{L}}$, only electron states are possible in the range $\Omega
_{3} $, whereas from the point of view of the right asymptotic area $S_{%
\mathrm{R}}$, only positron states are possible in this range. For the
detailed consideration of this fact in the framework of QED, see Sec. VII
of Ref.~\cite{x-case}.

In what follows, operators $a_{n}$ and $a_{n}^{\dag }$, and $b_{n}$ and $%
b_{n}^{\dag }$ denote operators of creation and annihilation of particles
(electrons) and antiparticles (positrons), respectively, for each range $%
\Omega _{i}$. Operators 
\begin{align}
& \ ^{-}a_{n}(\mathrm{in})=\ ^{-}a_{n},\ \ _{+}a_{n}(\mathrm{in})=\
_{+}a_{n},\   \notag \\
& \ _{-}b_{n}^{\dag }(\mathrm{in})=\ _{-}b_{n}^{\dag },\ \ ^{+}b_{n}(\mathrm{%
in})=\ ^{+}b_{n},  \label{op-in}
\end{align}%
and their conjugates correspond to the initial electrons and positrons,
while operators 
\begin{align}
& \ ^{+}a_{n}(\text{\textrm{out}})=\ ^{+}a_{n},\ \ _{-}a_{n}(\text{\textrm{%
out}})=\ _{-}a_{n},  \notag \\
& \ _{+}b_{n}(\text{\textrm{out}})=\ _{+}b_{n},\ \ ^{-}b_{n}(\text{\textrm{%
out}})=\ ^{-}b_{n},  \label{op-out}
\end{align}%
correspond to the final electrons and positrons.

The formalism developed in Ref.~\cite{x-case} is applicable to any
one-dimensional $x$-electric\ potential step as long as the condition (\ref%
{gen.00c}) is satisfied. When the solutions $^{\zeta }\psi _{n}(X)$\ and $%
_{\zeta }\psi _{n}(X)$\ of the corresponding Dirac equation with such a
potential can be found analytically in each spatial region $S_{\mathrm{L}}$, 
$S_{\mathrm{R}}$, and $S_{\mathrm{int}}$, it is possible to use border
conditions (gluing conditions) to calculate all main characteristics of
vacuum instability (the number of particles created from the vacuum, the
probability of the vacuum to remain a vacuum, etc.). The general procedure of
such calculations can be found in Refs.~\cite{x-case,L-field,x-exp}, where
several examples of exactly solvable cases are considered: the so-called $L$-constant field \cite{L-field}, the Sauter-like field \cite{x-case}, and the
peak electric field given by an exponential step \cite{x-exp}.

\section{Density operators with different initial conditions\label{initial}}

To obtain the density operators for the system under consideration, we
introduce the special generating functionals $R(J)$, which are given in Appendix A. Choosing the appropriate sources $J$, we are able to obtain the
explicit form for the density operators for different initial conditions.

\subsection{Initial vacuum state}

\noindent To obtain the density operator with an initial vacuum state, we set
all $J=0$ in $R(J)$, i.e., we set $J_{\pm ,n}^{(i)}=J_{n}^{(i)}=0$ in every
partial generating functional $R^{(i)}(J)$. In this case, the general
density operator with a vacuum initial state takes the form%
\begin{equation}
R(J=0)=\ \hat{\rho}_{v}=\otimes \prod\limits_{i=1}^{5}\hat{\rho}_{v}^{(i)},\
\ \hat{\rho}_{v}^{(i)}=\prod\limits_{n\in \Omega _{i}}\hat{\rho}%
_{v,n}^{(i)}\ ,  \label{vac.1}
\end{equation}%
where the one-mode partial density operators $\hat{\rho}_{n,v}^{(i)}$ are (in
terms of an \textrm{in} set of creation and annihilation operators\footnote{%
Here and in what follows colons $:\ldots:$ always denote the normal form
with respect to the creation and annihilation operators inside them.}) 
\begin{align}
& \hat{\rho}_{v,n}^{(1)}=\mathbf{:}\exp \left[ -\ _{+}a_{n}^{\dag }\
_{+}a_{n}\ -\ ^{-}a_{n}^{\dag }\ ^{-}a_{n}\ \right] \mathbf{:},\ \hat{\rho}%
_{v,n}^{(5)}=\mathbf{:}\exp \left[ -\ ^{+}b_{n}^{\dag }\ ^{+}b_{n}\ -\
_{-}b_{n}^{\dag }\ _{-}b_{n}\ \right] \mathbf{:},  \notag \\
& \hat{\rho}_{v,n}^{(3)}=\mathbf{:}\exp \left[ -\ ^{-}a_{n}^{\dag }\
^{-}a_{n}-\ _{-}b_{n}^{\dag }\ _{-}b_{n}\right] \mathbf{:},\ \hat{\rho}%
_{v,n}^{(2)}=\mathbf{:}\exp \left[ -a_{n}^{\dagger }\ a_{n}\right] \mathbf{:}%
,\ \hat{\rho}_{v,n}^{(4)}=\mathbf{:}\exp \left[ -b_{n}^{\dagger }\ b_{n}%
\right] \mathbf{:}.  \label{vac.2}
\end{align}%
Taking into account the well-known Berezin formula \cite{berez} 
\begin{equation}
\left\vert 0\right\rangle \left\langle 0\right\vert =\ \mathbf{:}\exp \left[
-\ a^{\dag }\ a\ \right] \mathbf{:},  \label{vac.2a}
\end{equation}%
one can see that the operators $\hat{\rho}_{v,n}^{(i)}$ are, in fact, partial
vacuum projectors for the initial particles:%
\begin{equation}
\hat{\rho}_{v,n}^{(i)}=\left\vert 0,\text{\textrm{in}}\right\rangle
_{n}^{(i)}\ _{n}^{(i)}\left\langle 0,\text{\textrm{in}}\right\vert ,\
i=1,3,5,\ \hat{\rho}_{v,n}^{(2,4)}=\left\vert 0\right\rangle _{n}^{(2,4)}\ \
_{n}^{(2,4)}\left\langle 0\right\vert .  \label{vac.3}
\end{equation}%
One can show that the differential numbers of initial electrons and
positrons [see Eqs. (\ref{op-in}) and (\ref{op-out}) for the reference] in
the state described by operator $\hat{\rho}_{v}$ vanish for all $n$,%
\begin{align}
& \text{\textrm{tr}}\hat{\rho}_{v}\ ^{-}a_{n}^{\dag }\ ^{-}a_{n}=\text{%
\textrm{tr}}\hat{\rho}_{v}\ _{+}a_{n}^{\dag }\ _{+}a_{n}=\text{\textrm{tr}}%
\hat{\rho}_{v}\ _{-}b_{n}^{\dag }\ _{-}b_{n}=\text{\textrm{tr}}\hat{\rho}%
_{v}\ ^{+}b_{n}^{\dag }\ ^{+}b_{n}=0,\ n\in \Omega _{1,3,5},  \notag \\
& \text{\textrm{tr}}\hat{\rho}_{v}\ a_{n}^{\dag }\ a_{n}=\text{\textrm{tr}}%
\hat{\rho}_{v}\ b_{n}^{\dag }\ b_{n}=0,\ n\in \Omega _{2,4}.  \label{vac.4}
\end{align}%
The mean differential numbers of final electrons and positrons are different
from zero in range $\Omega _{3}$. These numbers are equal to the number of
pairs created from vacuum,%
\begin{equation}
N_{n}^{a}=N_{n}^{b}=N_{n}^{\mathrm{cr}}=\text{\textrm{tr}}\hat{\rho}_{v}\
^{+}a_{n}^{\dag }\ ^{+}a_{n}=\left\vert g\left( _{-}|^{+}\right) \right\vert
^{-2},\ n\in \Omega _{3},  \label{vac.5}
\end{equation}%
where $g\left( _{-}|^{+}\right) $ are mutual decomposition coefficients of
the solutions $_{-}\psi _{n}(X)$ and $^{+}\psi _{n}(X)$ [see Eqs.~(\ref%
{gf.decomposition}) and (\ref{gf.dec.prop}].

\subsection{Initial thermal state}

\noindent Before writing the expressions for the density operator, we must
recall that we consider the situation when the electric field is not zero
only in the finite region $S_{\mathrm{int}}=\left( x_{\mathrm{L}},x_{\mathrm{%
R}}\right) $ situated between the planes $x=x_{\mathrm{L}}$ and $x=x_{%
\mathrm{R}}$. Outside of $S_{\mathrm{int}}$ for $x\in S_{\mathrm{L}%
}=(-\infty ,x_{\mathrm{L}}]$ and for $x\in S_{\mathrm{R}}=[x_{\mathrm{R}%
},\infty )$ particles are free (i.e., their movement is unbounded at least in
one direction). It should be noted that usually quantum field theory deals with physical
quantities that are presented by volume integrals on the hyperplane $t=$const. The main contribution to these integrals is from regions 
$S_{\mathrm{L}}$ and $S_{\mathrm{R}}$, where particles are free. This fact
allows one to obtain the explicit form of kinetic energies for all particles
(see details in Ref. \cite{x-case}) and is used in what follows.

To obtain the density operator with the initial thermal state, we need to set
the sources $J$ as
\begin{align}
& J_{\pm ,n}^{(i)}=e^{-E_{n\in \Omega _{i}}^{\pm }},\ E_{n}^{\pm }=\beta
\left( \varepsilon _{n}^{\pm }-\mu ^{\pm }\right) ,\ \beta =\Theta ^{-1},\
n\in \Omega _{1,3,5},\   \notag \\
& J_{n}^{(i)}=e^{-E_{n\in \Omega _{i}}},\ E_{n}=\beta \left( \varepsilon
_{n}-\mu \right) ,\ n\in \Omega _{2,4},\   \label{th.1}
\end{align}%
where $\varepsilon _{n}^{\pm }$ and $\varepsilon _{n}$ are the kinetic
energies of particles and antiparticles with quantum numbers $n$; $\mu ^{\pm
}$ and $\mu $ are the corresponding chemical potentials, and $\Theta $ is
the absolute temperature.\footnote{Here and later in the definition of von Neumann entropy we omit the Boltzmann
constant $k_{B}$ for the sake of convenience.} 
For the sake of simplicity, in what follows we will suppose that all chemical potentials for electrons
and positrons are equal. The density operator $\hat{\rho}_{\beta }$ can be
written as%
\begin{equation}
\hat{\rho}_{\beta }=\otimes \prod\limits_{i=1}^{5}\hat{\rho}_{\beta
}^{(i)},\ \ \hat{\rho}_{\beta }^{(i)}=\prod\limits_{n\in \Omega _{i}}\hat{%
\rho}_{\beta ,n}^{(i)},\   \label{th.2}
\end{equation}%
where the one-mode density operators $\hat{\rho}_{\beta ,n}^{(i)}$ have the form 
\begin{align}
& \ \hat{\rho}_{\beta ,n}^{(1)}=\left[ Z_{n}^{(1)}\right] ^{-1}\exp \left[
-\ _{+}a_{n}^{\dag }\ E_{n}^{+}\ _{+}a_{n}\ -\ ^{-}a_{n}^{\dag }\ E_{n}^{-}\
^{-}a_{n}\ \right] ,\   \notag \\
& \ \hat{\rho}_{\beta ,n}^{(5)}=\left[ Z_{n}^{(5)}\right] ^{-1}\exp \left[
-\ ^{+}b_{n}^{\dag }\ E_{n}^{+}\ ^{+}b_{n}-\ _{-}b_{n}^{\dag }\ E_{n}^{-}\
_{-}b_{n}\right] ,\ \   \notag \\
& \ \hat{\rho}_{\beta ,n}^{(3)}=\left[ Z_{n}^{(3)}\right] ^{-1}\exp \left[
-\ ^{-}a_{n}^{\dagger }\ E_{n}^{+}\ ^{-}a_{n}-\ _{-}b_{n}^{\dagger }\
E_{n}^{-}\ _{-}b_{n}\right] ,  \notag \\
& \ \hat{\rho}_{\beta ,n}^{(2)}=\left[ Z_{n}^{(2)}\right] ^{-1}\exp \left[
-\ a_{n}^{\dag }\ E_{n}\ a_{n}\right] \ \ \hat{\rho}_{\beta ,n}^{(4)}=\left[
Z_{n}^{(4)}\right] ^{-1}\exp \left[ -\ b_{n}^{\dag }\ E_{n}\ b_{n}\right] .
\label{th.3}
\end{align}%
The statistical sums $Z_{n}^{(i)}$ have the form 
\begin{equation}
Z_{n}^{(1,3,5)}=\left( 1+e^{-E_{n}^{+}}\right) \left(
1+e^{-E_{n}^{-}}\right) ,\ Z_{n}^{(2,4)}=\left( 1+e^{-E_{n}}\right) .
\label{th.4}
\end{equation}%
Note that the operators (\ref{th.3}) can also be presented as%
\begin{equation}
\hat{\rho}_{\beta ,n}^{(i)}=\left[ Z_{n}^{(i)}\right] ^{-1}\exp \left\{
-\beta \left[ \hat{H}_{n}^{(i)}-\mu \hat{N}_{n}^{(i)}\right] \right\} ,
\label{th.5}
\end{equation}%
where for $i=1,3,5$ we have 
\begin{align}
\hat{H}_{n}^{(i)}& =\left\{ 
\begin{array}{c}
\ _{+}a_{n}^{\dag }\ \varepsilon _{n}^{+}\ _{+}a_{n}+\ ^{-}a_{n}^{\dag }\
\varepsilon _{n}^{-}\ ^{-}a_{n},\ \ n\in \Omega _{1} \\ 
\ ^{-}a_{n}^{\dagger }\ \varepsilon _{n}^{+}\ ^{-}a_{n}+\ _{-}b_{n}^{\dagger
}\ \varepsilon _{n}^{-}\ _{-}b_{n},\ \ n\in \Omega _{3} \\ 
\ ^{+}b_{n}^{\dag }\ \varepsilon _{n}^{+}\ ^{+}b_{n}+\ _{-}b_{n}^{\dag }\
\varepsilon _{n}^{-}\ _{-}b_{n},\ \ n\in \Omega _{5}%
\end{array}%
\right. ,  \notag \\
\ \mu \hat{N}_{n}^{(i)}& =\left\{ 
\begin{array}{c}
\mu ^{+}\ _{+}a_{n}^{\dag }\ _{+}a_{n}+\mu ^{-}\ ^{-}a_{n}^{\dag }\
^{-}a_{n},\ \ n\in \Omega _{1} \\ 
\mu ^{+}\ ^{-}a_{n}^{\dagger }\ ^{-}a_{n}+\mu ^{-}\ _{-}b_{n}^{\dagger }\
_{-}b_{n},\ \ n\in \Omega _{3} \\ 
\mu ^{+}\ ^{+}b_{n}^{\dag }\ ^{+}b_{n}+\mu ^{-}\ _{-}b_{n}^{\dag }\
_{-}b_{n},\ \ n\in \Omega _{5}%
\end{array}%
\right. ,  \label{th.6}
\end{align}%
while for $i=2,4$ these operators take the form%
\begin{equation}
\hat{H}_{n}^{(i)}=\left\{ 
\begin{array}{c}
a_{n}^{\dag }\ \varepsilon _{n}\ a_{n},\ \ n\in \Omega _{2} \\ 
b_{n}^{\dag }\ \varepsilon _{n}\ b_{n},\ \ n\in \Omega _{4}%
\end{array}%
\right. ,\ \ \mu \hat{N}_{n}^{(i)}=\left\{ 
\begin{array}{c}
\mu \ a_{n}^{\dag }a_{n},\ \ n\in \Omega _{2} \\ 
\mu \ b_{n}^{\dag }b_{n},\ \ n\in \Omega _{4}%
\end{array}%
\right. .  \label{th.7}
\end{equation}%
The density operators (\ref{th.5}) in each range $\Omega _{i}$ are the
density operators of the grand canonical ensemble at temperature $\Theta $
and with chemical potentials $\mu ^{\pm }$, $\mu $. The differential mean
distributions $N_{n}^{(i)}$, calculated with the help of density matrices $%
\hat{\rho}_{\beta ,n}^{(i)}$, are well-known Fermi-Dirac and Bose-Einstein
distributions. For example, the differential number of initial electrons in the
range $\Omega _{3}$ can be found as%
\begin{equation}
N_{n,\beta ,-}^{(3)}(\text{\textrm{in}})=\text{\textrm{tr}}\hat{\rho}_{\beta
,n}^{(3)}\hat{N}_{n,\beta ,-}^{(3)}(\text{\textrm{in}})=\text{\textrm{tr}}%
\hat{\rho}_{\beta ,n}^{(3)}\ ^{-}a_{n}^{\dagger }\ ^{-}a_{n}=\left(
e^{E_{n}^{-}}+1\right) ^{-1},\ \ n\in \Omega _{3}.  \label{th.8}
\end{equation}%
Other differential distributions can be calculated in the same way using the
corresponding creation and annihilation operators and partial density
operators.

\section{Reduced density operators\label{reduced}}

\subsection{Reduced density operators for the electron and the positron
subsystems}

In the general case, the states of the system under consideration at the
final time instant contain both particles and antiparticles due to the pair
creation by external fields and the structure of the initial state. However,
we are often interested in physical quantities $F_{\pm}$ that describe only
electrons ($+$) or positrons ($-$) at the final state of the system. The
corresponding operators $\hat{F}_{\pm}$ are functions of either electron creation and annihilation operators $a$ and $a^{\dag}$ or positron ones $b$ and $b^{\dag}$. The mean values of these operators can be obtained from the
so-called reduced density operators $\hat{\rho}_{\pm}$, defined as reduced
traces of the general density matrix $\hat{\rho}$ over one of the subsystems
(the positron or electron one, respectively): 
\begin{equation}
\hat{\rho}_{\pm}=\text{\textrm{tr}}_{\mp}\hat{\rho},\ \ \hat{\rho}%
=\otimes\prod_{i=1}^{5}\hat{\rho}^{(i)},\ \hat{\rho}^{(i)}=\prod_{n\in
\Omega_{i}}\hat{\rho}_{n}^{(i)}.\   \label{rd.1}
\end{equation}
In the latter expression, the reduced traces \textrm{tr}$_{\mp}$ of the
operator $\hat{\rho}$ are defined as%
\begin{equation}
\text{\textrm{tr}}_{+}\hat{\rho}=\sum_{M=0}^{\infty}\sum_{\left\{ m\right\} }%
\frac{1}{M!}\left\langle \Psi_{\left\{ m\right\} _{M}}^{a}\right\vert \hat{%
\rho}\left\vert \Psi_{\left\{ m\right\} _{M}}^{a}\right\rangle ,\ \text{%
\textrm{tr}}_{-}\hat{\rho}=\sum_{M=0}^{\infty}\sum_{\left\{ m\right\} }\frac{%
1}{M!}\left\langle \Psi_{\left\{ m\right\} _{M}}^{b}\right\vert \hat{\rho}%
\left\vert \Psi_{\left\{ m\right\} _{M}}^{b}\right\rangle ,  \label{rd.2}
\end{equation}
where $\left\vert \Psi_{\left\{ m\right\} _{M}}^{a(b)}\right\rangle $ are
state vectors for electron (positron) states,%
\begin{equation}
\left\vert \Psi_{\left\{ m\right\} _{M}}^{a}\right\rangle =a_{m_{1}}^{\dag
}\ldots a_{m_{M}}^{\dag}\left\vert 0,\text{\textrm{out}}\right\rangle _{a},\
\ \left\vert \Psi_{\left\{ m\right\} _{M}}^{b}\right\rangle
=b_{m_{1}}^{\dag}\ldots b_{m_{M}}^{\dag}\left\vert 0,\text{\textrm{out}}%
\right\rangle _{b}.  \label{rd.3}
\end{equation}
Here $\left\vert 0,\text{\textrm{out}}\right\rangle _{a}$ and $\left\vert 0,%
\text{\textrm{out}}\right\rangle _{b}$ are partial electron and positron
vacua. Note that in the ranges $\Omega_{1}$ and $\Omega_{2}$ (where only electron states exist) $\left\vert 0,\text{\textrm{out}}\right\rangle
_{a}^{(1,2)}=\left\vert 0,\text{\textrm{out}}\right\rangle ^{(1,2)}$;
similarly to this, in the ranges $\Omega_{4}$ and $\Omega_{5}$ we have $%
\left\vert 0,\text{\textrm{out}}\right\rangle _{b}^{(4,5)}=\left\vert 0,%
\text{\textrm{out}}\right\rangle ^{(4,5)}$. In the Klein zone $\Omega_{3}$,
where both electron and positron states exist, the total vacuum is a product
of electron and positron partial vacua $\left\vert 0,\text{\textrm{out}}%
\right\rangle _{a}^{(3)}\otimes\left\vert 0,\text{\textrm{out}}\right\rangle
_{b}^{(3)}=\left\vert 0,\text{\textrm{out}}\right\rangle ^{(3)}$. Every
partial electron and positron vacuum can be presented in turn as a product
in quantum modes
\begin{equation}
\left\vert 0,\text{\textrm{out}}\right\rangle _{a}^{(i)}=\prod_{n\in\Omega
_{i}}\left\vert 0,\text{\textrm{out}}\right\rangle _{a,n}^{(i)},\ \
\left\vert 0,\text{\textrm{out}}\right\rangle
_{b}^{(i)}=\prod_{n\in\Omega_{i}}\left\vert 0,\text{\textrm{out}}%
\right\rangle _{b,n}^{(i)}.  \label{rd.3a}
\end{equation}
For this reason it is obvious enough that the reduced trace tr$_{+}$ completely traces out partial density operators $\hat{\rho}^{(1,2)}$
and leaves partial operators $\hat{\rho}^{(4,5)}$ unaffected. In the same
manner the reduced trace \textrm{tr}$_{-}$ traces out operators $\hat{\rho }%
^{(4,5)}$ and leaves $\hat{\rho}^{(1,2)}$ unchanged. Therefore, the reduced
density operators $\hat{\rho}_{\pm}$ can be presented as%
\begin{align}
& \hat{\rho}_{+}=\hat{\rho}^{(1)}\otimes\hat{\rho}^{(2)}\otimes\hat{\rho}%
_{+}^{(3)},\ \hat{\rho}_{+}^{(3)}=\text{\textrm{tr}}_{-}\ \hat{\rho}^{(3)}, 
\notag \\
& \hat{\rho}_{+}^{(3)}=\sum_{M=0}^{\infty}\sum_{\left\{ m\right\} }\left(
M!\right) ^{-1}\ _{b}^{(3)}\left\langle 0,\text{\textrm{out}}\right\vert \
_{+}b_{m_{M}}\ldots\ _{+}b_{m_{1}}\hat{\rho}^{(3)}\ _{+}b_{m_{1}}^{\dag
}\ldots\ _{+}b_{m_{M}}^{\dag}\left\vert 0,\text{\textrm{out}}\right\rangle
_{b}^{(3)},  \notag \\
& \hat{\rho}_{-}=\hat{\rho}^{(4)}\otimes\hat{\rho}^{(5)}\otimes\hat{\rho}%
_{-}^{(3)},\ \hat{\rho}_{-}^{(3)}=\text{\textrm{tr}}_{+}\ \hat{\rho}^{(3)}, 
\notag \\
& \hat{\rho}_{-}^{(3)}=\sum_{M=0}^{\infty}\sum_{\left\{ m\right\} }\left(
M!\right) ^{-1}\ _{a}^{(3)}\left\langle 0,\text{\textrm{out}}\right\vert \
^{+}a_{m_{M}}\ldots\ ^{+}a_{m_{1}}\hat{\rho}^{(3)}\ ^{+}a_{m_{1}}^{\dag
}\ldots\ ^{+}a_{m_{M}}^{\dag}\left\vert 0,\text{\textrm{out}}\right\rangle
_{a}^{(3)},  \label{rd.4}
\end{align}
The reduced density operators $\hat{\rho}_{\pm}^{(3)}$ can be obtained from
the reduced generating functionals $\underline{R}_{\pm}^{(3)}$,
\begin{equation}
\underline{R}_{\pm}^{(3)}=\text{\textrm{tr}}_{\mp}\underline{R}^{(3)},
\label{rd.5}
\end{equation}
where the partial traces are defined in the same way as in Eq.~(\ref{rd.4}).

Using the path-integral representation for traces (\ref{A.6}) in the
representation (\ref{gf.7}) for $\underline{R}^{(3)}$, we obtain that%
\begin{align}
& \underline{R}_{+}^{(3)}=\prod_{n\in \Omega _{3}}\left[ Z_{+,n}^{(3)}\right]
^{-1}:\exp \left\{ -\ ^{+}a_{n}^{\dag }\left[ 1-K_{+}(J)\right] \ ^{+}a_{n}%
\right\} :,  \notag \\
& \underline{R}_{-}^{(3)}=\prod_{n\in \Omega _{3}}\left[ Z_{-,n}^{(3)}\right]
^{-1}:\exp \left\{ -\ _{+}b_{n}^{\dag }\left[ 1-K_{-}(J)\right] \ _{+}b_{n}%
\right\} :,  \notag \\
& K_{\pm }(J)=D_{\pm }+C^{\dag }\left( 1+D_{\mp }\right) ^{-1}C,\ \left[
Z_{\pm ,n}^{(3)}\right] ^{-1}=\left\vert w_{n}(-|-)\right\vert ^{-2}\left(
1+AB\right) \left( 1+D_{\mp }\right) ,  \label{rd.6}
\end{align}%
where$\ A$, $B$, $C$, and $D_{\pm }$ are some functions of sources $J$ and
relative elementary amplitudes $w_{n}$ given by Eq. (\ref{gf.7}); $w_{n}(-|-) $ is the relative amplitude of the electron transition. The explicit
form of the elementary amplitudes $w_{n}$ is given by Eq. (\ref{omega}).
Choosing appropriate sources $J$ in the same manner as it was done in
Sec. \ref{initial}, we can obtain the corresponding partial density
operators $\hat{\rho}_{\pm }^{(3)}$ for different initial conditions.

\subsection{Measurement induced reduction}

We can also consider a reduction of density operators, which occurs due to
measurement of a physical quantity, namely, the number of final particles,
by some classical tool. This kind of reduction can also occur due to some
decoherence processes, such as collisions with some external sources (e.g.,
with impurities in graphene). For us, there is no difference which of
the mechanisms is implemented, so in what follows we talk about an
intermediate measurement by a classical tool as a source of the decoherence.

We study the measurement induced deformation of the density matrix for two
initial conditions, namely, when the initial state of the system is a
pure state and when the system initially is in a thermal equilibrium. Suppose that we are measuring
the number of final particles (electrons or positrons) $N$ in the state $%
\hat {\rho}$ of the system under consideration. According to von Neumann 
\cite{vN}, the density operator $\hat{\rho}$ after this measurement is
reduced to the operator $\hat{\rho}_{N}$ of the form%
\begin{equation}
\hat{\rho}_{N}=\sum_{\{s\}}W_{s}\hat{P}_{s},\ \hat{P}_{s}=\left\vert s,\text{%
\textrm{out}}\right\rangle \left\langle s,\text{\textrm{out}}\right\vert ,\
W_{s}=\left\langle s,\text{\textrm{out}}\right\vert \hat{\rho }\left\vert s,%
\text{\textrm{out}}\right\rangle ,  \label{rd.7}
\end{equation}
where $\left\vert s,\text{\textrm{out}}\right\rangle $ are eigenstates of
the operator $\hat{N}$ with the eigenvalues $s$ that represent the total
number of electrons and positrons in the state $\left\vert s,\text{\textrm{%
out}}\right\rangle $,%
\begin{align}
& \hat{N}(\text{\textrm{out}})|s,\text{\textrm{out}}\rangle =s|s,\text{%
\textrm{out}}\rangle,\ |s,\text{\textrm{out}}\rangle=\prod _{n\in\Omega_{1}}%
\left[ \text{ }^{+}a_{n}^{\dag}\right] ^{l_{n,1}}\left[ \text{ }%
_{-}a_{n}^{\dag}\right] ^{k_{n,1}}\prod_{n\in\Omega_{2}}\left[ \text{ }%
a_{n}^{\dag}\right] ^{l_{n,2}}  \notag \\
& \times\prod_{n\in\Omega_{4}}\left[ \text{ }b_{n}^{\dag}\right]
^{l_{n,4}}\prod_{n\in\Omega_{5}}\left[ \text{ }_{+}b_{n}^{\dag}\right]
^{l_{n,5}}\left[ \text{ }^{-}b_{n}^{\dag}\right] ^{k_{n,5}}\prod_{n\in%
\Omega_{3}}\left[ {\ }^{+}a_{n}^{\dag}\right] ^{l_{n,3}}\left[ {\ }%
_{+}b_{n}^{\dag }\right] ^{k_{n,3}}|0,\text{\textrm{out}}\rangle,  \notag \\
& s=\sum_{n\in\Omega_{1}}\left( l_{n,1}+k_{n,1}\right) +\sum_{n\in
\Omega_{2}}\left( l_{n,2}\right) +\sum_{n\in\Omega_{4}}\left( l_{n,4}\right)
+\sum_{n\in\Omega_{5}}\left( l_{n,5}+k_{n,5}\right)
+\sum_{n\in\Omega_{3}}\left( l_{n,3}+k_{n,3}\right) .  \label{rd.8}
\end{align}
Note that for Dirac particles $l_{n,i}$, $k_{n,i}=(0,1)$. It is convenient
to introduce partial density operators for each range $\Omega_{i}$: 
\begin{align}
& \hat{\rho}_{N}^{(i)}=\prod_{n\in\Omega_{i}}\hat{\rho}_{N,n}^{(i)},\ \hat{%
\rho}_{N,n}^{(i)}=\sum_{\{s_{i}\}}W_{s,n}^{(i)}\ \hat{P}_{s,n}^{(i)},\  
\notag \\
& \hat{P}_{s,n}^{(i)}=\ \left\vert s_{i},\text{\textrm{out}}\right\rangle
_{n}^{(i)}\ _{n}^{(i)}\left\langle s_{i},\text{\textrm{out}}\right\vert ,\
W_{s,n}^{(i)}=\ ^{(i)}\left\langle s_{i},\text{\textrm{out}}\right\vert \hat{%
\rho}^{(i)}\left\vert s_{i},\text{\textrm{out}}\right\rangle ^{(i)}.
\label{rd.12}
\end{align}
This way the general density operator of the system can be presented as%
\begin{equation}
\hat{\rho}_{N}=\otimes\prod_{i=1}^{5}\hat{\rho}_{N}^{(i)}.  \label{rd.12a}
\end{equation}
The state vectors $\left\vert s_{i},\text{\textrm{out}}\right\rangle
_{n}^{(i)}$ introduced in Eq. (\ref{rd.12}) are different for each range $%
\Omega_{i}$:%
\begin{align}
& \left\vert s_{1},\text{\textrm{out}}\right\rangle _{n}^{(1)}=\left[ \text{ 
}^{+}a_{n}^{\dag}\right] ^{l_{n,1}}\left[ \text{ }_{-}a_{n}^{\dag}\right]
^{k_{n,1}}\ \left\vert 0,\text{\textrm{out}}\right\rangle _{n}^{(1)},\
s_{1}=l_{n,1}+k_{n,1},  \notag \\
& \left\vert s_{3},\text{\textrm{out}}\right\rangle _{n}^{(3)}=\left[ {\ }%
^{+}a_{n}^{\dag}\right] ^{l_{n,3}}\left[ {\ }_{+}b_{n}^{\dag}\right]
^{k_{n,3}}\ \left\vert 0,\text{\textrm{out}}\right\rangle _{n}^{(3)},\
s_{3}=l_{n,3}+k_{n,3},  \notag \\
& \left\vert s_{5},\text{\textrm{out}}\right\rangle _{n}^{(5)}=\left[ \text{ 
}_{+}b_{n}^{\dag}\right] ^{l_{n,5}}\left[ \text{ }^{-}b_{n}^{\dag}\right]
^{k_{n,5}}\ \left\vert 0,\text{\textrm{out}}\right\rangle _{n}^{(5)},\
s_{5}=l_{n,5}+k_{n,5},  \notag \\
& \left\vert s_{2},\text{\textrm{out}}\right\rangle _{n}^{(2)}=\left[ \text{ 
}a_{n}^{\dag}\right] ^{l_{n,2}}\left\vert 0\right\rangle _{n}^{(2)},\
\left\vert s_{4},\text{\textrm{out}}\right\rangle _{n}^{(4)}=\left[ \text{ }%
b_{n}^{\dag}\right] ^{l_{n,4}}\left\vert 0\right\rangle _{n}^{(4)},\
s_{2/4}=l_{n,2/4}.  \label{rd.12b}
\end{align}
The sum of all eigenvalues is equal to the total number of particles in the
state $|s,$\textrm{out}$\rangle$, i.e., $\sum_{i=1}^{5}\sum_{n\in%
\Omega_{i}}s_{i}=s$. In what follows we also use the following notation for partial vacuum
projectors for the Klein zone:%
\begin{equation}
P_{v,n}^{(3)}(\mathrm{in})=|0,\mathrm{in}\rangle_{n}^{(3)}{}_{n}^{(3)}%
\langle0,\mathrm{in}|,\ P_{v,n}^{(i)}(\mathrm{out})=|0,\mathrm{out}\rangle
_{n}^{(i)}{}_{n}^{(i)}\langle0,\mathrm{out}|.  \label{rd.12c}
\end{equation}

\subsubsection{Initial vacuum state}

Vacuum states in ranges $\Omega _{1,2,4,5}$ remain in vacuum and the
measurement of the number of particles does not deform the partial density
operators with vacuum initial conditions\ in these ranges. It is easy to
show that in the Klein zone $\Omega _{3}$ the initial vacuum state evolves
as%
\begin{equation}
|0,\mathrm{in}\rangle _{n}^{(3)}=c_{\mathrm{v},n}\left[ 1-{\ }%
^{+}a_{n}^{\dag }\ w_{n}\left( +-|0\right) {\ }_{+}b_{n}^{\dag }\right] |0,%
\mathrm{out}\rangle _{n}^{(3)},\ \ c_{\mathrm{v},n}=\ _{n}^{(3)}\langle 0,%
\mathrm{out}|0,\mathrm{in}\rangle _{n}^{(3)},  \label{v.1}
\end{equation}%
where $w_{n}\left( +-|0\right) $ is a relative amplitude of pair production.
The corresponding partial density operator $\hat{\rho}_{v,n}^{(3)}$ with an
initial vacuum state can be written as%
\begin{equation}
\hat{\rho}_{v,n}^{(3)}=P_{v,n}^{(3)}(\mathrm{in})=\left\vert c_{\mathrm{v}%
,n}\right\vert ^{2}\left[ 1-{\ }^{+}a_{n}^{\dag }w_{n}\left( +-|0\right) {\ }%
_{+}b_{n}^{\dag }\right] {}P_{v,n}^{(3)}(\mathrm{out})\left[ 1-{\ }%
_{+}b_{n}w_{n}\left( +-|0\right) ^{\ast }{\ }^{+}a_{n}\right] .  \label{v.2}
\end{equation}%
Performing a reduction procedure (\ref{rd.7}), we obtain%
\begin{equation}
\hat{\rho}_{N,n}^{(3)}=\left\vert c_{\mathrm{v},n}\right\vert ^{2}\
P_{v,n}^{(3)}(\mathrm{out})+\left\vert c_{\mathrm{v},n}\right\vert
^{2}\left\vert w_{n}\left( +-|0\right) \right\vert ^{2}{\ }^{+}a_{n}^{\dag }{%
\ }_{+}b_{n}^{\dag }\ P_{v,n}^{(3)}(\mathrm{out}){\ }_{+}b_{n}{\ }^{+}a_{n}.
\label{v.3}
\end{equation}%
The first term of this expression corresponds to the situation where we find
a vacuum state after the measurement and the second one corresponds to the
situation where we find the state with an electron-positron pair. The coefficients $%
\left\vert c_{\mathrm{v},n}\right\vert ^{2}$ and $\left\vert c_{\mathrm{v}%
,n}\right\vert ^{2}\left\vert w_{n}\left( +-|0\right) \right\vert ^{2}$ are
classical probabilities for each of the outcomes.

\subsubsection{Initial thermal state}

We can consider the measurement-induced reduction for the thermal initial state
of the system. The partial density operators $\hat{\rho}_{n}^{(i)}$ are
obtained from the generating functionals $R^{(i)}$ by setting the sources $J$
as in Eq. (\ref{th.1}). The following are the nonvanishing weights $W^{(i)}$ from Eq. (\ref{rd.12}): the range $\Omega_{1}$,%
\begin{align}
& W_{1,n}^{(1)}=\ _{n}^{(1)}\left\langle 0,\text{\textrm{out}}\right\vert 
\hat{\rho}_{\beta,n}^{(1)}\left\vert 0,\text{\textrm{out}}\right\rangle
_{n}^{(1)}=\left[ Z_{n}^{(1)}\right] ^{-1},\   \notag \\
& W_{2,n}^{(1)}=\ _{n}^{(1)}\left\langle 0,\text{\textrm{out}}\right\vert \
^{+}a_{n}\ \hat{\rho}_{\beta,n}^{(1)}\ ^{+}a_{n}^{\dag}\left\vert 0,\text{%
\textrm{out}}\right\rangle _{n}^{(1)}=\ \left[ Z_{n}^{(1)}\right] ^{-1}%
\tilde{C}_{++},\ \tilde{C}_{++}=1+C_{++},  \notag \\
& W_{3,n}^{(1)}=\ _{n}^{(1)}\left\langle 0,\text{\textrm{out}}\right\vert
_{-}a_{n}\ \hat{\rho}_{\beta,n}^{(1)}\ _{-}a_{n}^{\dag}\left\vert 0,\text{%
\textrm{out}}\right\rangle _{n}^{(1)}=\ \left[ Z_{n}^{(1)}\right] ^{-1}%
\tilde{C}_{--},\ \tilde{C}_{--}=1+C_{--},  \notag \\
& W_{4,n}^{(1)}=\ _{n}^{(1)}\left\langle 0,\text{\textrm{out}}\right\vert \
^{+}a_{n}\ _{-}a_{n}\ \hat{\rho}_{\beta,n}^{(1)}\ _{-}a_{n}^{\dag}\
^{+}a_{n}^{\dag}\left\vert 0,\text{\textrm{out}}\right\rangle _{n}^{(1)}=\ %
\left[ Z_{n}^{(1)}\right] ^{-1}\ \left[ \tilde{C}_{++}\tilde {C}%
_{--}-C_{+-}C_{-+}\right] ;  \label{rd.14d}
\end{align}
the range $\Omega_{5}$,%
\begin{align}
& W_{1}^{(5)}=\ _{n}^{(5)}\left\langle 0,\text{\textrm{out}}\right\vert \hat{%
\rho}_{\beta,n}^{(5)}\left\vert 0,\text{\textrm{out}}\right\rangle
_{n}^{(5)}=\left[ Z_{n}^{(5)}\right] ^{-1},\   \notag \\
& W_{2}^{(5)}=\ _{n}^{(5)}\left\langle 0,\text{\textrm{out}}\right\vert \
_{+}b_{n}\ \hat{\rho}_{\beta,n}^{(5)}\ _{+}b_{n}^{\dag}\left\vert 0,\text{%
\textrm{out}}\right\rangle _{n}^{(5)}=\ \left[ Z_{n}^{(5)}\right] ^{-1}%
\tilde{D}_{++},\ \tilde{D}_{++}=1+D_{++},  \notag \\
& W_{3}^{(5)}=\ _{n}^{(5)}\left\langle 0,\text{\textrm{out}}\right\vert
^{-}b_{n}\ \hat{\rho}_{\beta,n}^{(5)}\ ^{-}b_{n}^{\dag}\left\vert 0,\text{%
\textrm{out}}\right\rangle _{n}^{(5)}=\ \left[ Z_{n}^{(5)}\right] ^{-1}%
\tilde{D}_{--},\ \tilde{D}_{--}=1+D_{--},  \notag \\
& W_{4}^{(5)}=\ _{n}^{(5)}\left\langle 0,\text{\textrm{out}}\right\vert \
_{+}b_{n}\ ^{-}b_{n}\ \hat{\rho}_{\beta,n}^{(5)}\ \ ^{-}b_{n}^{\dag}\
_{+}b_{n}^{\dag}\left\vert 0,\text{\textrm{out}}\right\rangle _{n}^{(5)}=\ %
\left[ Z_{n}^{(5)}\right] ^{-1}\left[ \tilde{D}_{++}\tilde{D}%
_{--}-D_{+-}D_{-+}\right] ;  \label{rd.14e}
\end{align}
the range $\Omega_{3}$,%
\begin{align}
& W_{1}^{(3)}=\ _{n}^{(3)}\left\langle 0,\text{\textrm{out}}\right\vert \hat{%
\rho}_{\beta,n}^{(3)}\left\vert 0,\text{\textrm{out}}\right\rangle
_{n}^{(3)}=\tilde{Z}_{n}^{(3)},\ \ \tilde{Z}_{n}^{(3)}=\left[ Z_{n}^{(3)}%
\right] ^{-1}\left\vert c_{\mathrm{v},n}\right\vert ^{2}\left( 1+AB\right)
,\   \notag \\
& W_{2}^{(3)}=\ _{n}^{(3)}\left\langle 0,\text{\textrm{out}}\right\vert \
^{+}a_{n}\ \hat{\rho}_{\beta,n}^{(3)}\ ^{+}a_{n}^{\dag}\left\vert 0,\text{%
\textrm{out}}\right\rangle _{n}^{(3)}=\ \tilde{Z}_{n}^{(3)}D_{+},  \notag \\
& W_{3}^{(3)}=\ _{n}^{(3)}\left\langle 0,\text{\textrm{out}}\right\vert
_{+}b_{n}\ \hat{\rho}_{\beta,n}^{(3)}\ _{+}b_{n}^{\dag}\left\vert 0,\text{%
\textrm{out}}\right\rangle _{n}^{(3)}=\ \tilde{Z}_{n}^{(3)}D_{-},  \notag \\
& W_{4}^{(3)}=\ _{n}^{(3)}\left\langle 0,\text{\textrm{out}}\right\vert \
^{+}a_{n}\ _{+}b_{n}\ \hat{\rho}_{\beta,n}^{(3)}\ _{+}b_{n}^{\dag}\
^{+}a_{n}^{\dag}\left\vert 0,\text{\textrm{out}}\right\rangle _{n}^{(3)}=\ 
\tilde{Z}_{n}^{(3)}\left( D_{+}D_{-}+C^{\dag}C\right) .  \label{rd.14f}
\end{align}

\subsection{Spatial reduction (left and right subsystems)}

The $x$-electric potential steps provide the spatial separation of the whole
system in two subsystems, the left subsystem and right subsystem, i.e., final
particles to the left of potential step and final particles to the right of
potential step. It is easy to imagine a situation when we are interested in
measuring physical values only in left and right asymptotic areas. For
example, we can suppose that measuring tools are situated only to the left
of the potential step. In this case the general density operator must be
averaged (reduced) over all unavailable states of the final right particles.

From the general theory \cite{x-case} we know the following. In the range $%
\Omega_{3}$ all the electrons (initial and final) are located on the left
side of the potential step and all the positrons are on the right side. Therefore, one can see that a reduction over the left and right
subsystems in the range $\Omega_{3}$ coincides with a reduction over electron and
positron subsystems, respectively, i.e., 
\begin{equation}
\hat{\rho}_{\text{$\mathrm{left}$}}^{(3)}=\hat{\rho}_{+}^{(3)},\ \ \hat{\rho}%
_{\text{$\mathrm{right}$}}^{(3)}=\hat{\rho}_{-}^{(3)}.  \label{rd.10a}
\end{equation}
In the range $\Omega_{2}$ there are only electrons on the left side of potential
step, so $\hat{\rho}_{\text{\textrm{left}}}^{(2)}=\hat{\rho}^{(2)}$.
Similarly, there are only right positrons in the range $\Omega_{4}$; therefore $%
\hat{\rho}_{\text{\textrm{right}}}^{(4)}=\hat{\rho}^{(4)}$.

The range $\Omega_{1}$ contains left and right electrons; the range $\Omega_{5}$
contains left and right positrons. In these ranges we need to consider the
reduction of partial generating functionals over the left or right final
particles. Let us start with the range $\Omega_{1}$. It is convenient to use the
expression for the generating functional $\underline{R}^{(1)}$ given by Eq. (\ref{gf.3g}) to calculate the partial (right and left) trace over states
with right electrons, i.e., states constructed with creation and annihilation
operators $^{+}a^{\dag}$ and $^{+}a$, or $\ _{-}a^{\dag}$ and $_{-}a$, thus
creating spatially reduced partial generating functionals as
\begin{align}
& \underline{R}_{\text{$\mathrm{left}$}}^{(1)}=\mathrm{tr}_{\text{$\mathrm{%
right}$}}^{(1)}\underline{R}^{(1)},\ \ \underline{R}_{\text{$\mathrm{right}$}%
}^{(1)}=\text{\textrm{tr}}_{\text{$\mathrm{left}$}}^{(1)}\underline{R}^{(1)},
\notag \\
& \mathrm{tr}_{\text{$\mathrm{right}$}}\hat{A}=\sum_{M=0}^{\infty}\sum_{%
\left\{ m\right\} }\left( M!\right) ^{-1}\left\langle \Psi_{\left\{
m\right\} _{M}}^{\text{$\mathrm{right}$}}\right\vert A\left\vert
\Psi_{\left\{ m\right\} _{M}}^{\text{$\mathrm{right}$}}\right\rangle , 
\notag \\
& \text{\textrm{tr}}_{\text{$\mathrm{left}$}}\hat{A}=\sum_{M=0}^{\infty}%
\sum_{\left\{ m\right\} }\left( M!\right) ^{-1}\left\langle \Psi_{\left\{
m\right\} _{M}}^{\text{$\mathrm{left}$}}\right\vert A\left\vert
\Psi_{\left\{ m\right\} _{M}}^{\text{$\mathrm{left}$}}\right\rangle ,
\label{rd.7a}
\end{align}
where $\left\vert \Psi_{\left\{ m\right\} _{M}}^{\text{$\mathrm{right}$}%
}\right\rangle $ and $\left\vert \Psi_{\left\{ m\right\} _{M}}^{\text{$%
\mathrm{left}$}}\right\rangle $ are state vectors for right and left
electrons, respectively,
\begin{align}
& \left\vert \Psi_{\left\{ m\right\} _{M}}^{\text{$\mathrm{right}$}%
}\right\rangle =\ ^{+}a_{m_{1}}^{\dag}\ldots\ ^{+}a_{m_{M}}^{\dag}\left\vert
0,\text{\textrm{out}}\right\rangle _{\text{$\mathrm{right}$}}^{(1)},  \notag
\\
& \left\vert \Psi_{\left\{ m\right\} _{M}}^{\text{$\mathrm{left}$}%
}\right\rangle =\ _{-}a_{m_{1}}^{\dag}\ldots\ _{-}a_{m_{M}}^{\dag}\left\vert
0,\text{\textrm{out}}\right\rangle _{\text{$\mathrm{left}$}}^{(1)},  \notag
\\
& \left\vert 0,\text{\textrm{out}}\right\rangle _{\text{$\mathrm{right}$}%
}^{(1)}\otimes\left\vert 0,\text{\textrm{out}}\right\rangle _{\text{$\mathrm{%
left}$}}^{(1)}=\left\vert 0,\text{\textrm{out}}\right\rangle _{\ }^{(1)}.
\label{rd.8a}
\end{align}
Note that partial left and right electron vacua can be factorized in quantum
modes $n$ and therefore generating functionals (\ref{rd.7a}) can be
factorized as well. Taking this into account and calculating the trace, we
obtain%
\begin{align}
& \underline{R}_{\text{$\mathrm{left},n$}}^{(1)}=\left( 1+\tilde{C}%
_{++}\right) \mathbf{:}\exp\left\{ \ _{-}a_{n}^{\dag}\ L_{-}\
_{-}a_{n}\right\} \mathbf{:},\ L_{-}=C_{--}-C_{+-}\left( 1+\tilde{C}%
_{++}\right) ^{-1}C_{-+},  \notag \\
& \underline{R}_{\text{$\mathrm{right},n$}}^{(1)}=\left( 1+\tilde{C}%
_{--}\right) \mathbf{:}\exp\left\{ \ ^{+}a_{n}^{\dag}\ L_{+}\
^{+}a_{n}\right\} \mathbf{:},\ L_{+}=C_{++}-C_{+-}\left( 1+\tilde{C}%
_{--}\right) ^{-1}C_{-+}.  \label{rd.9a}
\end{align}
Similar results can be obtained for the range $\Omega_{5}$:%
\begin{align}
& \underline{R}_{\text{$\mathrm{right},n$}}^{(5)}=\left( 1+\tilde{D}%
_{--}\right) \mathbf{:}\exp\left\{ \ _{+}b_{n}^{\dag}\ K_{+}\
_{+}b_{n}^{\dag}\right\} \mathbf{:},\ K_{+}=D_{++}-D_{+-}\left( 1+\tilde{D}%
_{--}\right) ^{-1}D_{-+},  \notag \\
& \underline{R}_{\text{$\mathrm{left},n$}}^{(5)}=\left( 1+\tilde{D}%
_{++}\right) \mathbf{:}\exp\left\{ \ ^{-}b_{n}^{\dag}\ K_{-}\
^{-}b_{n}\right\} \mathbf{:},\ K_{-}=D_{--}-D_{+-}\left( 1+\tilde{D}%
_{++}\right) ^{-1}D_{-+}.  \label{rd.9b}
\end{align}
Now setting the sources $J$ as discussed in Sec. \ref{initial},
one can obtain the density operators $\hat{\rho}_{\text{$\mathrm{left (right)}$%
}}^{(1,5)}$ with different initial conditions.

\section{Entropy of reduced density operators\label{Neumann}}

It is known that the electron-positron pair is always produced in an
entangled state. The process of pair production does not change the entropy of
the whole system, since its evolution is unitary. It is easy, however, to
imagine a situation in which only a certain quantum subsystem is available
for measurements; in this case, we must reduce the total density matrix over
the states of the inaccessible subsystem. Reduction over one of the quantum
subsystems makes part of the information unavailable \cite{NieCh00}. Thus, the
reduced density operators we have introduced in the preceding section always
describe mixed states even when the initial state of the system is pure.
This means that the entropy of particular subsystems can change as the
subsystems become entangled due to pair creation. This entropy change can be
used as a measure of information loss due to reduction or as a measure of
entanglement between those subsystems. In what follows we show that the
above-mentioned entropy change due to pair production can be quantified via
the corresponding numbers of final particles, which depend on the choice of
initial state and on the type of reduction, or, in other words, on the choice of
quantum subsystem. By controlling the strength and duration of the electric
field applied to the system, we have the theoretical ability to change the
average number of pairs generated by the field, and thus, their
entanglement. Another important task is to understand how a measurement of a
physical value (for example, a measurement of the final mean number of
particles) can change the entropy of the system. To address all these
questions we calculate the entropy, corresponding to different reductions of
the general density matrix for different initial conditions.

As a measure of information loss due to reduction we use the von Neumann
entropy, defined as 
\begin{equation}
S(\hat{\rho})=-\text{\textrm{tr}}\hat{\rho}\ln \hat{\rho}\ .  \label{en.1}
\end{equation}

\subsection{Entropy corresponding to reduction over the subsystems of
electrons and positrons}

First, we calculate von Neumann
entropy for reduced density matrices (\ref{rd.4}), 
\begin{equation}
S(\hat{\rho}_{\pm })=-\text{\textrm{tr}}\hat{\rho}_{\pm }\ln \hat{\rho}_{\pm
}\text{\ },  \label{en.3}
\end{equation}%
where $\mathrm{tr}$ denotes the full trace of the operator,\ \textrm{tr}$%
\hat{A}=$\textrm{tr}$_{-}$\textrm{tr}$_{+}\hat{A}$.

Using the definitions (\ref{rd.4}), let us transform the operator $\ln \hat{%
\rho}_{\pm}$ as follows 
\begin{equation}
\ln\hat{\rho}_{+}=\ln\hat{\rho}^{(1)}+\ln\hat{\rho}^{(2)}+\ln\hat{\rho}%
_{+}^{(3)},\ \ \ln\hat{\rho}_{-}=\ln\hat{\rho}^{(5)}+\ln\hat{\rho}^{(4)}+\ln%
\hat{\rho}_{-}^{(3)}.  \label{en.4}
\end{equation}
Due to the fact that partial density matrices $\hat{\rho}^{(i)}$ and $\hat{%
\rho}_{\pm}^{(3)}$ are normalized (\textrm{tr}$\hat{\rho}^{(i)}=$ \textrm{tr}%
$\hat{\rho}_{\pm}^{(3)}=1$), it is easy to show that Eq. (\ref{en.3})
transforms into the sum of entropies
\begin{equation}
S(\hat{\rho}_{+})=S(\hat{\rho}^{(1)})+S(\hat{\rho}^{(2)})+S(\hat{\rho}%
_{+}^{(3)}),\ \ S(\hat{\rho}_{-})=S(\hat{\rho}^{(5)})+S(\hat{\rho}^{(4)})+S(%
\hat{\rho}_{-}^{(3)}).  \label{en.5}
\end{equation}
We recall that in each range $\Omega_{i}$ the partial density operators 
$\hat{\rho}^{(i)}$ and $\hat{\rho}_{\pm}^{(3)}$ can be factorized in quantum
modes $n$ and each one-mode operator is also normalized. This allows us to
further simplify the expressions (\ref{en.5}) and write%
\begin{align}
& S(\hat{\rho}^{(i)})=\sum_{n\in\Omega_{i}}S(\hat{\rho}_{n}^{(i)})=-\sum_{n%
\in\Omega_{i}}\text{\textrm{tr}}\hat{\rho}_{n}^{(i)}\ln\hat{\rho }%
_{n}^{(i)},\ \ i=1,2,4,5,  \notag \\
& S(\hat{\rho}_{\pm}^{(3)})=\sum_{n\in\Omega_{3}}S(\hat{\rho}%
_{\pm,n}^{(3)})=-\sum_{n\in\Omega_{3}}\text{\textrm{tr}}\hat{\rho}%
_{\pm,n}^{(3)}\ln\hat{\rho}_{\pm,n}^{(3)}.  \label{en.6}
\end{align}
Now one can calculate the entropy for the
density operators with different initial conditions.

\subsubsection{Initial vacuum state}

\noindent In this case partial density operators $\hat{\rho}_{n}^{(i)}$, $i=1,2,4,5$, are given by Eq. (\ref{vac.2}). It is easy to see that the
corresponding entropies vanish, i.e., $S(\hat{\rho}_{n}^{(1,2,4,5)})=0$. The
entropies of density operators $\hat{\rho}_{\pm ,n}^{(3)}$ are equal and can
be calculated \cite{unitarity} to have the form%
\begin{equation}
S(\hat{\rho}_{\pm ,n}^{(3)})=-\left[ \left( 1-N_{n}^{\text{$\mathrm{cr}$}%
}\right) \ln \left( 1-N_{n}^{\text{$\mathrm{cr}$}}\right) +N_{n}^{\text{$%
\mathrm{cr}$}}\ln N_{n}^{\text{$\mathrm{cr}$}}\right] ,  \label{en.7}
\end{equation}%
where $N_{n}^{\text{\textrm{cr}}}$ is the mean differential number of pairs
created from vacuum by the electric field (\ref{vac.5}). The total entropy can
be found as a sum over all quantum numbers in $\Omega _{3}$,%
\begin{equation}
S\left( \hat{\rho}_{\pm }^{(3)}\right) =-\sum_{n\in \Omega _{3}}\left[
\left( 1-N_{n}^{\text{$\mathrm{cr}$}}\right) \ln \left( 1-N_{n}^{\text{$%
\mathrm{cr}$}}\right) +N_{n}^{\text{$\mathrm{cr}$}}\ln N_{n}^{\text{$\mathrm{%
cr}$}}\right] .  \label{en.7a}
\end{equation}

\subsubsection{Initial thermal state}

\noindent The entropies corresponding to partial density operators $\hat{%
\rho }_{\beta,n}^{(i)}$ are 
\begin{subequations}
\begin{align}
& S(\hat{\rho}_{\beta,n}^{(1,5)})=-\sum_{\zeta=\pm}\left\{ \left[
1-N_{n,\beta,\zeta}^{(1,5)}(\text{\textrm{in}})\right] \ln\left[
1-N_{n,\beta,\zeta}^{(1,5)}(\text{\textrm{in}})\right] +N_{n,\beta,\zeta
}^{(1,5)}(\text{\textrm{in}})\ln N_{n,\beta,\zeta}^{(1,5)}(\text{\textrm{in}}%
)\right\} ,  \label{en.8a}   \\
& S(\hat{\rho}_{\beta,n}^{(2,4)})=-\left\{ \left[ 1-N_{n,\beta}^{(2,4)}(\text{%
\textrm{in}})\right] \ln\left[ 1-N_{n,\beta}^{(2,4)}(\text{\textrm{in}}%
)\right] +N_{n,\beta}^{(2,4)}(\text{\textrm{in}})\ln N_{n,\beta}^{(2,4)}(%
\text{\textrm{in}})\right\} ,\   \label{en.8b} \\
& S(\hat{\rho}_{\beta,\pm,n}^{(3)})=-\left\{ \left[ 1-N_{n,\beta,\pm}^{(3)}(%
\text{\textrm{out}})\right] \ln\left[ 1-N_{n,\beta,\pm}^{(3)}(\text{\textrm{%
out}})\right] +N_{n,\beta,\pm}^{(3)}(\text{\textrm{out}})\ln
N_{n,\beta,\pm}^{(3)}(\text{\textrm{out}})\right\} ,\   \label{en.8}
\end{align}
\end{subequations}
where the mean differential numbers of particles from the Eqs. (\ref{en.8a}) and (\ref{en.8b})
are given by 
\begin{equation}
N_{n,\beta,\zeta}^{(1)}(\text{\textrm{in}})=\left(
e^{E_{n}^{\zeta}}+1\right) ^{-1},\ \zeta=\pm,\ n\in\Omega_{1,5},\ \
N_{n,\beta}(\text{\textrm{in}})=\left( e^{E_{n}}+1\right) ^{-1},\
n\in\Omega_{2,4},  \label{en.9}
\end{equation}
and $N_{n,\beta,\pm}^{(3)}($\textrm{out}$)$ are the differential mean
numbers of final electrons ($+$) and positrons ($-$) in the range $\Omega_{3}$,%
\begin{align}
& N_{n,\beta,+}^{(3)}(\mathrm{out})=\text{\textrm{tr}}\hat{\rho}_{+,n,\beta
}^{(3)}\ ^{+}a_{n}^{\dag}\ ^{+}a_{n}=N_{n}^{\text{$\mathrm{cr}$}}\left(
1-N_{n,\beta,-}^{(3)}(\text{\textrm{in}})\right) +\left( 1-N_{n}^{\text{$%
\mathrm{cr}$}}\right) N_{n,\beta,+}^{(3)}(\text{\textrm{in}}),  \notag \\
& N_{n,\beta,-}^{(3)}(\mathrm{out})=\text{\textrm{tr}}\hat{\rho}_{-,n,\beta
}^{(3)}\ \ _{+}b_{n}^{\dag}\ _{+}b_{n}=N_{n}^{\text{$\mathrm{cr}$}}\left(
1-N_{n,\beta,+}^{(3)}(\text{\textrm{in}})\right) +\left( 1-N_{n}^{\text{$%
\mathrm{cr}$}}\right) N_{n,\beta,-}^{(3)}(\text{\textrm{in}}).  \label{en.10}
\end{align}
The differential mean numbers $N_{n,\beta,\pm}^{(3)}($\textrm{in}$)$ in 
Eq. (\ref{en.10}) can be calculated similarly to Eq. (\ref{th.8}) using the
corresponding creation and annihilation operators.

\subsection{ Entropy corresponding to measurement induced reduction}

The measurement reduced density operators $\hat{\rho}_{N}^{(i)}$ with
different initial conditions are given by Eq. (\ref{rd.12}). Similarly to
the preceding section, it is easy to show that the von Neumann entropy can be
presented as the sum over quantum modes $n$ of partial entropies,%
\begin{equation}
S(\hat{\rho}_{N}^{(i)})=\sum_{n\in\Omega_{i}}S(\hat{\rho}_{N,n}^{(i)})=-%
\sum_{n\in\Omega_{i}}\text{\textrm{tr}}\hat{\rho}_{N,n}^{(i)}\ln\hat{\rho }%
_{N,n}^{(i)}.  \label{mr.1}
\end{equation}
Therefore, to obtain the total entropy it is sufficient to calculate only
the entropies $S(\hat{\rho}_{N,n}^{(i)})$ corresponding to partial density
operators $\hat{\rho}_{N,n}^{(i)}$ and then perform the summation over all
quantum numbers $n\in\Omega_{i}$.

\subsubsection{Initial vacuum state}

Let us calculate von Neumann entropy corresponding to density operator $\hat{%
\rho}_{N,n}^{(3)}$. We can show that the entropy for this case takes the form 
\begin{align}
& S(\hat{\rho}_{N,n}^{(3)})=-\left[ \,|c_{\mathrm{v},n}|^{2}\ln |c_{\mathrm{v%
},n}|^{2}+|c_{\mathrm{v},n}|^{2}|w_{n}\left( +-|0\right) |^{2}\ln|c_{\mathrm{%
v},n}|^{2}\!|w_{n}\left( +-|0\right) |^{2}\right] ,  \notag \\
& |c_{\mathrm{v},n}|^{2}=1-N_{n}^{\text{$\mathrm{cr}$}},\ \ \left\vert
w_{n}(+-|0)\right\vert ^{2}=N_{n}^{\text{$\mathrm{cr}$}}\left( 1-N_{n}^{%
\text{$\mathrm{cr}$}}\right) ^{-1},  \label{mr.2}
\end{align}
which leads us to the result 
\begin{equation}
S(\hat{\rho}_{N,n}^{(3)})=-\left[ \left( 1-N_{n}^{\text{$\mathrm{cr}$}%
}\right) \ln\left( 1-N_{n}^{\text{$\mathrm{cr}$}}\right) +N_{n}^{\text{$%
\mathrm{cr}$}}\ln N_{n}^{\text{$\mathrm{cr}$}}\right] .  \label{mr.3}
\end{equation}

\subsubsection{Initial thermal state}

For this case the density operators $\hat{\rho}_{N,n}^{(1)}$, $\hat{\rho}%
_{N,n}^{(3)}$, and $\hat{\rho}_{N,n}^{(5)}$ have the form (\ref{rd.12})
with the weights $W$ given by Eq. (\ref{rd.14d}) in $\Omega _{1}$, 
by Eq. (\ref{rd.14e}) in $\Omega _{5}$, and by Eq. (\ref{rd.14f}) in $\Omega _{3}$.
It can be shown that entropies $S(\hat{\rho}_{N,n}^{(i)})$ take the form 
\begin{equation}
S(\hat{\rho}_{N,n}^{(i)})=-\sum_{l=1}^{4}W_{l}^{(i)}\ln W_{l}^{(i)}.
\label{mr.8}
\end{equation}%
Sources $J$, given in Eq. (\ref{th.1}) for the case of the thermal initial
state are connected to differential mean numbers of initial particles by the relation
\begin{equation}
J_{\pm ,n}^{(i)}=e^{-E_{n\in \Omega _{i}}^{\pm }}=N_{n,\beta ,\pm }^{(i)}(%
\text{\textrm{in}})\left[ 1-N_{n,\beta ,\pm }^{(i)}(\text{\textrm{in}}%
)\right] ^{-1}.  \label{mr.9}
\end{equation}%
Using then Eqs. (\ref{RT})-(\ref{CtoRT}) and (\ref{gf.7}),
it is possible to present weights $W_{l}^{(i)}$ via differential mean
numbers of initial particles, reflection $\left\vert R_{n}\right\vert ^{2}$
and transition $\left\vert T_{n}\right\vert ^{2}$ probabilities, and the number
of particles created from vacuum $N_{n}^{\mathrm{cr}}$.

\subsection{Entropy corresponding to spatial reduction (left and right)}

We can also calculate von Neumann entropy for the left and right reduced
density operators, found in Eqs. (\ref{rd.9a}) and (\ref{rd.9b}). For the
reduced generating functionals from $\Omega_{1}$ this entropy has the form%
\textsf{\ }%
\begin{align}
& S(R_{\text{$\mathrm{left}$}}^{(1)})=-\text{\textrm{tr}}R_{\text{$\mathrm{%
left}$}}^{(1)}\mathrm{\ln}R_{\text{$\mathrm{left}$}}^{(1)}=\sum_{n\in%
\Omega_{1}}\left[ \ln Z_{n}^{(1)}-\ln\left( 1+\tilde {C}_{++}\right) -\
N_{n,-}^{(1)}(\mathrm{out})\ln\left( 1+L_{-}\right) \right] ,  \notag \\
& S(R_{\text{$\mathrm{right}$}}^{(1)})=-\text{\textrm{tr}}R_{\text{$\mathrm{%
right}$}}^{(1)}\mathrm{\ln}R_{\text{$\mathrm{right}$}}^{(1)}=\sum_{n\in%
\Omega_{1}}\left[ \ln Z_{n}^{(1)}-\ln\left( 1+\tilde {C}_{--}\right) -\
N_{n,+}^{(1)}(\mathrm{out})\ln\left( 1+L_{+}\right) \right] ,  \label{en.11}
\end{align}
where $N_{n,-}^{(1)}(\mathrm{out})$ and$\ N_{n,+}^{(1)}(\mathrm{out})$ are
the differential mean numbers of left and right final electrons in $%
\Omega_{1}$, 
\begin{align}
& N_{n,-}^{(1)}(\mathrm{out})=\text{\textrm{tr}}R_{\text{$\mathrm{left}$}%
}^{(1)}\ _{-}a_{n}^{\dag}\ _{-}a_{n}=\left[ Z_{n}^{(1)}\right] ^{-1}\left[
(1+\tilde{C}_{++})\tilde{C}_{--}-C_{+-}C_{-+}\right] ,\   \notag \\
& N_{n,+}^{(1)}(\mathrm{out})=\text{\textrm{tr}}R_{\text{$\mathrm{right}$}%
}^{(1)}\ \ ^{+}a_{n}^{\dag}\ ^{+}a_{n}=\left[ Z_{n}^{(1)}\right] ^{-1}\left[
(1+\tilde{C}_{--})\tilde{C}_{++}-C_{+-}C_{-+}\right] ,  \label{en.12}
\end{align}
Using the fact that reduced generating functionals $R_{\text{\textrm{left}}%
}^{(1)}\ $are normalized, \textrm{tr}$R_{\text{\textrm{left}}}^{(1)}=1$, one
can show that the following relations hold true:%
\begin{equation}
 1+L_{-}=\frac{N_{n,-}^{(1)}(\mathrm{out})}{1-N_{n,-}^{(1)}(%
\mathrm{out})},\  1+L_{+} =\frac{N_{n,+}^{(1)}(\mathrm{out})}{%
1-N_{n,+}^{(1)}(\mathrm{out})},\ \frac{1+\tilde{C}_{\pm\pm}}{Z_{n}^{(1)}}%
=\left( 2+L_{\mp}\right) ^{-1}.\   \label{en.12a}
\end{equation}
Using the these expressions, we can represent Eq. (\ref{en.11}) as%
\begin{align}
& S(R_{\text{$\mathrm{left}$}}^{(1)})=-\sum_{n\in\Omega_{1}}\left\{ \left[
1-N_{n,-}^{(1)}(\mathrm{out})\right] \ln\left[ 1-N_{n,-}^{(1)}(\mathrm{out}%
)\right] +N_{n,-}^{(1)}(\mathrm{out})\ln N_{n,-}^{(1)}(\mathrm{out})\right\} ,
\notag \\
& S(R_{\text{$\mathrm{right}$}}^{(1)})=-\sum_{n\in\Omega_{1}}\left\{ \left[
1-N_{n,+}^{(1)}(\mathrm{out})\right] \ln\left[ 1-N_{n,+}^{(1)}(\mathrm{out}%
)\right] +N_{n,+}^{(1)}(\mathrm{out})\ln N_{n,+}^{(1)}(\mathrm{out})\right\} .
\label{en.12b}
\end{align}

For the reduced generating functionals from $\Omega_{5}$ the result reads%
\begin{align}
& S(R_{\text{$\mathrm{left}$}}^{(5)})=-\text{\textrm{tr}}R_{\text{$\mathrm{%
left}$}}^{(5)}\mathrm{\ln}R_{\text{$\mathrm{left}$}}^{(5)}=\sum_{n\in%
\Omega_{5}}\left[ \ln Z_{n}^{(5)}-\ln\left( 1+\tilde {D}_{++}\right)
-N_{n,+}^{(5)}(\mathrm{out})\ln\left( 1+K_{+}\right) \right] ,  \notag \\
& S(R_{\text{$\mathrm{right}$}}^{(5)})=-\text{\textrm{tr}}R_{\text{$\mathrm{%
right}$}}^{(5)}\mathrm{\ln}R_{\text{$\mathrm{right}$}}^{(5)}=\sum_{n\in%
\Omega_{5}}\left[ \ln Z_{n}^{(5)}-\ln\left( 1+\tilde {D}_{--}\right)
-N_{n,-}^{(5)}(\mathrm{out})\ln\left( 1+K_{-}\right) \right] ,  \label{en.13}
\end{align}
where $N_{n,+}^{(5)}(\mathrm{out})$ and $N_{n,-}^{(5)}(\mathrm{out})$ are
the differential mean numbers of final positrons, 
\begin{align}
& N_{n,+}^{(5)}(\mathrm{out})=\text{\textrm{tr}}R_{\text{$\mathrm{left}$}%
}^{(5)}\ _{+}b_{n}^{\dag}\ _{+}b_{n}=\left[ Z_{n}^{(5)}\right] ^{-1}\left[
(1+\tilde{D}_{++})\tilde{D}_{--}-D_{+-}D_{-+}\right] ,\   \notag \\
& N_{n,-}^{(5)}(\mathrm{out})=\text{\textrm{tr}}R_{\text{$\mathrm{right}$}%
}^{(5)}\ ^{-}b_{n}^{\dag}\ ^{-}b_{n}=\left[ Z_{n}^{(5)}\right] ^{-1}\left[
(1+\tilde{D}_{--})\tilde{D}_{++}-D_{+-}D_{-+}\right] .  \label{en.14}
\end{align}
The entropies (\ref{en.13}) in terms of mean differential numbers (\ref%
{en.14}) take the form%
\begin{align}
& S(R_{\text{$\mathrm{left}$}}^{(5)})=-\sum_{n\in\Omega_{5}}\left\{ \left[
1-N_{n,+}^{(5)}(\mathrm{out})\right] \ln\left[ 1-N_{n,+}^{(5)}(\mathrm{out}%
)\right] +N_{n,+}^{(5)}(\mathrm{out})\ln N_{n,+}^{(5)}(\mathrm{out})\right\} ,
\notag \\
& S(R_{\text{$\mathrm{right}$}}^{(5)})=-\sum_{n\in\Omega_{5}}\left\{ \left[
1-N_{n,-}^{(5)}(\mathrm{out})\right] \ln\left[ 1-N_{n,-}^{(5)}(\mathrm{out}%
)\right] +N_{n,-}^{(5)}(\mathrm{out})\ln N_{n,-}^{(5)}(\mathrm{out})\right\} .
\label{en.14a}
\end{align}

\subsection{Loss of information due to electron-positron reduction in
L-constant field}

Here we illustrate some of the obtained formulas by considering the
deformation of the quantum vacuum between two infinite capacitor plates
separated by a finite distance $L$. Several aspects of particle creation by the
constant electric field between such plates (this field is also called the $L$%
-constant electric field) were studied in Ref. \cite{L-field}. The latter
field is a particular case of the $x$-electric potential step. Thus, we consider
the $L$-constant electric field in $d=D+1$ dimensions. We choose $\mathbf{E}(x)=E^{i}$, $\left(i=1,...,D\right)$,
$E^{1}=E_{x}(x)$, $E^{2,...,D}=0$, and
\begin{equation*}
E_{x}(x)=\left\{ 
\begin{array}{l}
0,\ x\in(-\infty,-L/2] \\ 
E=\mathrm{const}>0,\ x\in(-L/2,L/2) \\ 
0,\ x\in\lbrack L/2,\infty)%
\end{array}
\right. .
\end{equation*}
We consider a particular case with a sufficiently large length $L$ between
the capacitor plates, 
\begin{equation}
\sqrt{eE}L\gg\max\left\{ 1,E_{c}/E\right\} .  \label{L-large}
\end{equation}
Here $E_{c}=m^{2}/e$\ is the critical Schwinger field. We conditionally call
this approximation as large work approximation when $\Delta U=eEL\gg2m$. 
Such kind of an $x$-electric step represents a regularization for a
constant uniform electric field and is suitable for imitating a
small-gradient field.{\large \ }

\subsubsection{Initial vacuum state}

Let us calculate von Neumann entropy corresponding to subsystems of
electrons and positrons created from vacuum in $\Omega_{3}$. 
The leading asymptotic contributions to the differential and total number of particles
created from the vacuum in the large work approximation have the
form \cite{L-field}%
\begin{equation}
N_{n}^{\text{$\mathrm{cr}$}}\approx \exp \left[ -\pi \frac{\pi _{\bot }^{2}}{%
eE}\right] ,\ \ N^{\text{$\mathrm{cr}$}}\approx \frac{J_{(d)}TV(eE)^{d/2}}{%
(2\pi )^{d-1}}\exp \left( -\pi \frac{E_{c}}{E}\right) ,  \label{6.9}
\end{equation}%
where $V=$ $LV_{\bot }$ is the volume inside the capacitor (the volume
occupied by the electric field, $L$ is the distance between capacitor
plates, and $V_{\bot }$ is the transversal volume of capacitor), $%
J_{(d)}=2^{[d/2]-1}$ is a spin summation factor,\footnote{%
Here $[\ldots ]$ denotes the integer part of the expression.} and $e>0$ is
an absolute value of electron charge.

Let us estimate the information loss of the reduced electron and positron
subsystems, which can be calculated as entropies (\ref{en.7a}) of these
states. Performing summation over quantum modes $n$ (for the details of this
operation see Refs. \cite{L-field,unitarity}), we obtain the
expression 
\begin{equation}
S(\hat{\rho}_{\pm }^{(3)})\approx \frac{J_{(d)}TV(eE)^{d/2}}{(2\pi )^{d-1}}%
\exp \left( -\pi \frac{E_{c}}{E}\right) A\left( d,E_{c}/E\right) \text{ }\;%
\mathrm{if}\;d>2,  \label{6.14}
\end{equation}%
where the factor $A\left( d,E_{c}/E\right) $ has the form%
\begin{align}
& \ A\left( d,E_{c}/E\right) =\left[ \left( \pi E_{c}/E+d/2-1\right) \right.
\notag \\
& \left. +\sum_{l=1}^{\infty }\left[ l^{-d/2}-l^{-1}(l+1)^{(2-d)/2}\exp
\left( -\pi E_{c}/E\right) \right] \exp \left( -\pi \left( l-1\right)
E_{c}/E\right) \right] .  \label{6.15}
\end{align}%
Comparing Eqs. (\ref{6.14}) and (\ref{6.9}), one can see that the entropy is
proportional to the total number of particles created, i.e.,%
\begin{equation}
S(\hat{\rho}_{\pm }^{(3)})\approx N^{\text{$\mathrm{cr}$}}A\left(
d,E_{c}/E\right) .  \label{6.16}
\end{equation}%
The result coincides with that obtained for the $T$-constant
electric field in Ref.~\cite{stat}, i.e., we reproduce exactly the same
expression for the entropy of the electron-positron subsystem for the vacuum
initial state. This result shows that despite the fact that $L$-constant and 
$T$-constant electric fields are physically distinct, they can be considered
as two different regularizations of the uniform constant electric field in
the limit $T$, $L\rightarrow \infty $.

\subsubsection{Initial thermal state}

Here we only consider the Klein zone $\Omega_{3}$ as well, as for the case
of the electron-positron subsystem reduction the density operators of the other
quantum ranges $\Omega_{i}$ either are completely traced out and do not
contribute to the von Neumann entropy, or are undisturbed by the reduction and
therefore their initial entropy does not change after the reduction.

We take the Fermi distributions as those of the initial particles. They depend on
particle energy and are given by Eq. (\ref{th.1}). In the Klein zone these
distributions have the form%
\begin{equation}
N_{n,\beta,\pm}^{(3)}(\text{\textrm{in}})=\left\{ \exp\left[ \beta\left(
\varepsilon_{n}^{\pm}-\mu^{\pm}\right) \right] +1\right\} ^{-1}.\ 
\label{6.20}
\end{equation}
At any given $p_{\bot}$ the available quantum numbers $p_{0}$ in the Klein zone
for the $L$-constant field are restricted by the definition of the Klein zone 
\cite{x-case}
\begin{equation}
U_{\mathrm{L}}+\pi_{\bot}\leq p_{0}\leq U_{\mathrm{R}}-\pi_{\bot },\ \ U_{%
\mathrm{R}}=-U_{\mathrm{L}}=\Delta U/2=eEL/2  \label{6.21}
\end{equation}
such that%
\begin{equation}
\varepsilon_{n}^{\pm}=\pm p_{0}+\frac{\Delta U}{2}\left[ 1-N_{n}^{\text{$%
\mathrm{cr}$}}\right] .  \label{gav1}
\end{equation}
Here $U_{\mathrm{L}}=-eA_{0}(x\rightarrow-\infty)$ and $U_{\mathrm{R}%
}=-eA_{0}(x\rightarrow+\infty)$ are the left and right asymptotic potential
energies, respectively.

Let us analyze Eq. (\ref{6.20}) for initial electrons. The number $N_{n}^{%
\text{$\mathrm{cr}$}}$ is even with respect to the change $p_{0}%
\rightarrow-p_{0} $ and has the form (\ref{6.9}) for the large range if $%
\left\vert p_{0}\right\vert$, $\pi_{\bot}\ll\Delta U/2$. At the left (right)
edge of the Klein zone asymptotic longitudinal momenta $\left\vert p^{%
\mathrm{L}}\right\vert$ ($\left\vert p^{\mathrm{R}}\right\vert $),%
\begin{equation*}
\left\vert p^{\mathrm{L(R)}}\right\vert =\sqrt{\left[ \pm p_{0}+\Delta U/2%
\right] ^{2}-\pi_{\bot}^{2}},
\end{equation*}
tends to zero and one of the following limits holds true: $N_{n}^{\text{$%
\mathrm{cr}$}}\sim\left\vert p^{\mathrm{L}}\right\vert /\sqrt {eE}%
\rightarrow0$ or $N_{n}^{\text{$\mathrm{cr}$}}\sim\left\vert p^{\mathrm{R}%
}\right\vert /\sqrt{eE}\rightarrow0$, respectively. We see that kinetic
energies $\varepsilon_{n}^{\pm}$ tend to the minimum, given by transversal
energy $\varepsilon_{n}^{\pm}\rightarrow\pi_{\bot}$. Therefore, it is more
likely to find a particle with a lower kinetic energy $\sim\pi_{\bot}$,
just as one would expect.

For further analysis it is convenient to rewrite the expressions (\ref{en.10})
for the final differential number of electrons and positrons as%
\begin{align}
& N_{n,\beta,+}^{(3)}(\mathrm{out})=N_{n,\beta,+}^{(3)}(\text{\textrm{in}}%
)+N_{n}^{\text{$\mathrm{cr}$}}\left[ 1-N_{n,\beta,-}^{(3)}(\text{\textrm{in}}%
)-N_{n,\beta,+}^{(3)}(\text{\textrm{in}})\right] ,  \notag \\
& N_{n,\beta,-}^{(3)}(\mathrm{out})=N_{n,\beta,-}^{(3)}(\text{\textrm{in}}%
)+N_{n}^{\text{$\mathrm{cr}$}}\left[ 1-N_{n,\beta,-}^{(3)}(\text{\textrm{in}}%
)-N_{n,\beta,+}^{(3)}(\text{\textrm{in}})\right] .  \label{6.22}
\end{align}
Note that if $\mu^{+}=\mu^{-}=\mu$, the sum $N_{n,\beta,-}^{(3)}(\mathrm{in}%
)+N_{n,\beta,+}^{(3)}(\mathrm{in})$ is even with respect to the change $%
p_{0}\rightarrow-p_{0}$. Further consideration can be easily extended to
the case when, for example, $N_{n,\beta,+}^{(3)}($\textrm{in}$)=0$ or $%
N_{n,\beta,-}^{(3)}($\textrm{in}$)=0$, i.e., when only one type of initial
particle is present. We can sum these expression over quantum numbers $%
n\in\Omega_{3}$ as 
\begin{equation}
N_{\beta,\pm}^{(3)}(\mathrm{out})=\sum_{n\in\Omega_{3}}N_{n,\beta,\pm}^{(3)}(%
\mathrm{out})=\frac{J_{(d)}TV_{\bot}}{(2\pi)^{d-1}}\int_{p_{\bot},p_{0}\in%
\Omega_{3}}d^{d-2}p_{\bot}dp_{0}N_{n,\beta,\pm}^{(3)}(\mathrm{out}).
\label{6.23}
\end{equation}

It was shown in Ref. \cite{L-field} that a leading contribution to $N_{n}^{%
\text{$\mathrm{cr}$}}$, given by Eq. (\ref{6.9}), comes from the inner subrange $D$, defined as%
\begin{align}
& \frac{\pi_{\bot}}{\sqrt{eE}}<K_{\bot},\ \ \left\vert p_{0}\right\vert /%
\sqrt{eE}<\sqrt{eE}L/2-K,  \notag \\
& \sqrt{eE}L/2\gg K\gg K_{\bot}^{2}\gg\max\left\{ 1,m^{2}/eE\right\} .
\label{6.26}
\end{align}
For the second terms of Eqs. (\ref{6.22}) $N_{n}^{\text{$\mathrm{cr}$}}$ acts
as a cutoff factor, so we can integrate over subrange $D$ only. Note that
for quantum modes $n^{\prime}$ where $N_{n^{\prime}}^{\text{$\mathrm{cr}$}}$
is small enough, i.e., the number of particles created is small enough,
distributions $N_{n,\beta,\pm}^{(3)}(\mathrm{out})$ are only slightly differ
from initial distributions $N_{n,\beta,\pm}^{(3)}($\textrm{in}$)$. In this
situation, the corresponding entropy will almost coincide with the initial
entropy of each subsystem,%
\begin{equation}
S(\hat{\rho}_{\pm,n^{\prime}}^{(3)})\approx-\left\{ \left[ 1-N_{n^{\prime
},\beta,\pm}^{(3)}(\text{\textrm{in}})\right] \ln\left[ 1-N_{n^{\prime
},\beta,\pm}^{(3)}(\text{\textrm{in}})\right]
+N_{n^{\prime},\beta,\pm}^{(3)}(\text{\textrm{in}})\ln
N_{n^{\prime},\beta,\pm}^{(3)}(\text{\textrm{in}})\right\} .  \label{6.27}
\end{equation}
To calculate the impact of a pair creation we can rewrite Eq. (\ref{6.23}) as%
\begin{equation}
N_{\beta,\pm}^{(3)}(\mathrm{out})\approx\frac{J_{(d)}TV_{\bot}}{(2\pi)^{d-1}}%
\int_{p_{\bot},p_{0}\in D}d^{d-2}p_{\bot}dp_{0}N_{n,\beta,\pm}^{(3)}(\mathrm{%
out}).  \label{6.28}
\end{equation}
Let us consider, for example, the case $N_{n}^{\text{$\mathrm{cr}$}}\ll1$.
Taking the relation (\ref{gav1}) and integrating $N_{n,\beta,\pm}^{(3)}($in$)$
over $p_{0}$, we obtain that the leading term is%
\begin{align}
& \int_{D}dp_{0}N_{n,\beta,+}^{(3)}(\text{\textrm{in}})=N_{\bot,\beta,\pm
}^{(3)}(\text{\textrm{in}}),  \notag \\
& N_{\bot,\beta,\pm}^{(3)}(\text{\textrm{in}})\approx\frac{1}{\beta}\ln 
\frac{1+\exp\left[ -\beta\left( \sqrt{eE}K-\mu\right) \right] }{1+\exp\left[
-\beta\left( eEL-\mu\right) \right] }  \label{6.24}
\end{align}
In particular, for low temperature and not very large $\mu$, $%
\sqrt{eE}K\gg\mu$, we have $\beta\left( eEL-\mu\right) \gg\beta\left( \sqrt{%
eE}K-\mu\right) \gg1$ and then%
\begin{equation}
N_{\bot,\beta,\pm}^{(3)}(\text{\textrm{in}})\approx\frac{1}{\beta}%
\ln\left\vert 1+\exp\left[ -\beta\left( \sqrt{eE}K-\mu\right) \right]
\right\vert \approx\frac{1}{\beta}\exp\left[ -\beta\left( \sqrt{eE}%
K-\mu\right) \right] .  \label{gav2}
\end{equation}
For high temperature, $1\gg\beta\left( eEL-\mu\right) \gg$ $\beta\left( 
\sqrt{eE}K-\mu\right) $, 
\begin{equation}
N_{\bot,\beta,\pm}^{(3)}(\text{\textrm{in}})\approx\frac{1}{2}\left( eEL-%
\sqrt{eE}K\right) .  \label{gav3}
\end{equation}

Integrating it over the transversal momentum, we get%
\begin{equation}
N_{\beta,\pm}^{(3)}(\text{\textrm{in}})\approx\frac{J_{(d)}TV_{\bot}}{%
(2\pi)^{d-1}}(2K_{\bot})^{d-2}\left( eE\right) ^{d/2-1}N_{\bot,\beta,\pm
}^{(3)}(\text{\textrm{in}})\approx\frac{J_{(d)}TV\left( eE\right) ^{d/2}}{%
2(2\pi)^{d-1}}(2K_{\bot})^{d-2}.  \label{6.29}
\end{equation}

The second terms of (\ref{6.22}) can be integrated in a similar way,%
\begin{align}
& \ \int_{D}d^{d-2}p_{\bot}dp_{0}N_{n}^{\text{$\mathrm{cr}$}}\left[
1-N_{n,\beta,-}^{(3)}(\text{\textrm{in}})-N_{n,\beta,+}^{(3)}(\text{\textrm{%
in}})\right]  \notag \\
& \ =\int_{D}d^{d-2}p_{\bot}N_{n}^{\text{$\mathrm{cr}$}}\left[ eEL-2\sqrt{eE}%
K-N_{\bot,\beta,+}^{(3)}(\text{\textrm{in}})-N_{\bot,\beta ,-}^{(3)}(\text{%
\textrm{in}})\right] .  \label{6.30}
\end{align}
Note that for high temperature, the second terms of (\ref{6.22})\ vanish.
This means that for this particular case $N_{\beta,\pm}^{(3)}(\mathrm{out}%
)\approx N_{\beta,\pm}^{(3)}(\mathrm{in})$, and the starting entropy of the
system does not change significantly due to pair creation and subsequent
reduction over one of the subsystems.

The corresponding entropy for the general case is not difficult to write, 
\begin{equation}
S\left( \hat{\rho}_{\pm }^{(3)}\right) =-\sum_{n\in D}\left\{ \left[
1-N_{n,\beta ,\pm }^{(3)}(\text{\textrm{out}})\right] \ln \left[
1-N_{n,\beta ,\pm }^{(3)}(\text{\textrm{out}})\right] +N_{n,\beta ,\pm
}^{(3)}(\text{\textrm{out}})\ln N_{n,\beta ,\pm }^{(3)}(\text{\textrm{out}})%
\right\} ,  \label{6.31}
\end{equation}%
where the summation over the quantum numbers can be done in the same manner
as in Eq. (\ref{6.23}). However, unlike the case of the vacuum initial state,
the expression (\ref{6.31}) is complicated. To obtain further results from it one
must utilize numerical calculations with definite parameters of a particular
system configuration: temperature $\Theta =\beta ^{-1}$, field strength $E$,
and capacitor length $L$.

We can see that for very low temperatures, $\beta \rightarrow \infty $ and $%
N_{n,\beta ,\pm }^{(3)}($\textrm{in}$)\rightarrow 0$, we reproduce the
result obtained for the vacuum initial condition. Similar to the case of a
vacuum initial state, Eq.~(\ref{6.31}) reproduces the one that can be
obtained for the case of a $T$-constant electric field in the first order of
magnitude, supporting the conclusion that both fields can be considering the
two different regularizations of the uniform constant electric field in the
limit $T$, $L\rightarrow \infty $.

We also note that there is the following difference when considering the thermal 
initial state for an $x$-electric potential step. Unlike the case of time-dependent electric
fields, the initial left and right subsystems are spatially separated and
may in principle have different temperatures.

\section{Concluding remarks}

In this work, we have considered the deformation of different initial states
by constant nonuniform electric fields and statistical properties of the
resulting states. We have introduced a special generating functional
that allow us to construct density operators for different initial
conditions. In graphene and similar materials any electric field can be
considered as critical due to the fact that charge carriers are
massless. Because of this a significant number of carrier pairs is produced.
Possible dissipative processes lead to a loss of coherence of the states
arising from vacuum, and it becomes necessary to study the statistical
properties of the state generated by the field. For this reason, we
considered two cases of initial states of the system other than vacuum: the case
when the system was initially in thermodynamic equilibrium at an absolute
temperature $\Theta =\beta ^{-1}$ and the case of a pure state with a
certain number of particles with fixed quantum numbers. In the framework of
QED with $x$-electric potential steps, we have to introduce five partial
generating functionals for each range of quantum numbers $\Omega $. To
simplify further calculations, we construct the normal form of these
generating functionals in terms of creation and annihilation operators
corresponding to final particles. Setting appropriate sources in these
generating functionals, we obtain density operators for different initial
states of the system: the vacuum state, pure states with a definite number of
particles with fixed quantum numbers, and the thermal initial state. We also note
that it is formally possible to construct the generating functional for a
system with different initial conditions in different areas of quantum
numbers $\Omega $. For example, choosing $J_{\pm
,n}^{(3)}=J_{n}^{(2)}=J_{n}^{(4)}=0$ and $J_{\pm ,n}^{(1)}=J_{\pm
,n}^{(5)}=e^{-E_{n}^{\pm }}$, we can construct the following density operator:%
\begin{equation}
\hat{\rho}_{\text{mix}}=\prod_{n\in \Omega _{1}}\hat{\rho}_{\beta
,n}^{(1)}\otimes \prod_{n\in \Omega _{2}}\hat{\rho}_{v,n}^{(2)}\otimes
\prod_{n\in \Omega _{3}}\hat{\rho}_{v,n}^{(3)}\otimes \prod_{n\in \Omega
_{4}}\hat{\rho}_{v,n}^{(4)}\otimes \prod_{n\in \Omega _{5}}\hat{\rho}_{\beta
,n}^{(5)}.  \label{d.1}
\end{equation}%
This density operator corresponds to the case when there are particles with
quantum numbers $n$ from ranges $\Omega _{1}$ and $\Omega _{5}$ in thermal
equilibrium at the initial time instant, but there are no particles that
belong to the Klein zone $\Omega _{3}$ (i.e., the initial state in this range was
the vacuum state) as well as in the ranges $\Omega _{2}$ and $\Omega _{4}$.
Moreover, the functionals $R^{(i)}(J)$ permit factorization in quantum modes 
$n$ and each of those modes evolve separately. This fact allows one to
assemble the general density operator as a product of partial operators $%
R_{n}^{(i)}(J),$ setting their initial conditions individually for each mode 
$n$. Sometimes there are situations when only part of the system is
available for observation; in this case we need to construct reduced density
operators that describe this available part only. Another possible scenario
for reduction is a measurement with the classical tool, which causes
decoherence and deforms the general density operator of the system. We
considered three types of reduction: the reduction over electron and
positron subsystems, the reduction due to the measurement of number of final
particles, and the spatial reduction over left or right final particles. We
note that the latter kind of reduction is of interest when considering the
type of fields that are concentrated in restricted space areas. We can
compare the situation at hand to the case of QED with $t$-electric
potential steps \cite{Gitman}: For time-dependent uniform fields spatial
reduction always coincides with reduction over electron or positron
subsystems, as formulation of the problem suggests that the field occupies
the entire space. Therefore, for a field acting for a sufficiently long time
period all the final electrons, regardless of their initial state, move in the
direction of the field (and all the final positrons move in the opposite
direction). The same can be said about electron-positron pairs created from
vacuum by an external field. For this reason the electron subsystem always
coincides with the left spatial subsystem, and the positron subsystem coincides
with the right spatial subsystem in uniform time-dependent electric fields.
However, for $x$-potential electric steps \cite{x-case} the electric field
is restricted in a finite area of space. Thus, there exist initial particles
in ranges $\Omega _{1}$ and/or $\Omega _{5}$, which can go through the
potential barrier and end up at the opposite side of the potential barrier as
free final particles. Taking that into account, we can conclude that the
spatial reduction is different from electron-positron subsystem reduction in the
general case. However, this difference exists only when there are initial
particles in ranges $\Omega _{1}$ and $\Omega _{5}$. When there are no
initial particles in these ranges, e.g., for the case of the initial vacuum
state, spatial reduction coincides with electron-positron subsystem
reduction. We have constructed reduced density operators corresponding to
each of the three types of reduction. We have calculated von Neumann entropy for
the reduced density operators. Using the so-called $L$-constant field as an
example, we have shown that for the reduced density operators of electron
and positron subsystems this entropy is proportional to the total number of
pairs created. Comparing the result obtained for the $L$-constant
field to that obtained for the $T$-constant electric field in Ref.~\cite{stat}, 
we reproduced exactly the same expressions for the entropy of the
electron-positron subsystem. This result shows, that despite the
fact that $L$-constant and $T$-constant electric fields are physically
distinct, they can be considered as two different regularizations of the
uniform constant electric field in the limit $T$, $L\rightarrow \infty$.

\section{Acknowledgements}
S.P.G. and D.M.G. acknowledge support from Tomsk State University
Competitiveness Improvement Program and partial support from the Russian
Foundation for Basic Research, under Project No. 18-02-00149. D.M.G. was 
also supported by Funda{\c{c}}{\~{a}}o de Amparo {\`{a}}
Pesquisa do Estado de S{\~{a}}o Paulo (FAPESP) through Grant No. 16/03319-6, and by Conselho Nacional
de Desenvolvimento Cient{\'{i}}fico e Tecnol{\'{o}}gico (CNPq). The work of A.A.S. was
supported by FAPESP through Grant 17/05734-3.

\section*{Appendix A}

\appendix\setcounter{equation}{0} \renewcommand{\theequation}{A%
\arabic{equation}} 

In this appendix we present some more results regarding QED with $x$%
-electric potential steps, which may be useful for the reader. The operators $%
\hat{\Psi}_{i}(X)$ for each particular range $\Omega _{i}$ can be decomposed
using the specific sets of solutions of the Dirac equation with quantum
numbers $n\in \Omega _{i}$. These decompositions have the form%
\begin{align}
& \hat{\Psi}_{1}(X)=\sum_{n\in \Omega _{1}}\mathcal{M}_{n}^{-1/2}\left[ \
_{+}a_{n}(\mathrm{in})\ _{+}\psi _{n}(X)+\ ^{-}a_{n}(\mathrm{in})\ ^{-}\psi
_{n}(X)\right]   \notag \\
& =\sum_{n\in \Omega _{1}}\mathcal{M}_{n}^{-1/2}\left[ \ ^{+}a_{n}(\text{%
\textrm{out}})\ ^{+}\psi _{n}(X)+\ _{-}a_{n}(\text{\textrm{out}})\ _{-}\psi
_{n}(X)\right] ,  \notag \\
& \hat{\Psi}_{3}(X)=\sum_{n\in \Omega _{3}}\mathcal{M}_{n}^{-1/2}\left[ \
^{-}a_{n}(\mathrm{in})\ ^{-}\psi _{n}(X)+\ _{-}b_{n}^{\dag }(\mathrm{in})\
_{-}\psi _{n}(X)\right]   \notag \\
& =\sum_{n\in \Omega _{3}}\mathcal{M}_{n}^{-1/2}\left[ \ ^{+}a_{n}(\text{%
\textrm{out}})\ ^{+}\psi _{n}(X)+\ _{+}b^{\dag }(\text{\textrm{out}})\
_{+}\psi _{n}(X)\right] ,  \notag \\
& \hat{\Psi}_{5}(X)=\sum_{n\in \Omega _{5}}\mathcal{M}_{n}^{-1/2}\left[ \
^{+}b_{n}^{\dag }(\mathrm{in})\ ^{+}\psi _{n}(X)+\ _{-}b_{n}^{\dag }(\mathrm{%
in})\ _{-}\psi _{n}(X)\right]   \notag \\
& =\sum_{n\in \Omega _{5}}\mathcal{M}_{n}^{-1/2}\left[ \ _{+}b_{n}^{\dag }(%
\text{\textrm{out}})\ _{+}\psi _{n}(X)+\ ^{-}b_{n}^{\dag }(\text{\textrm{out}%
})\ ^{-}\psi _{n}(X)\right]   \label{gen.1}
\end{align}%
in the ranges $\Omega _{i}$, $i=1,3,5$, and%
\begin{equation}
\hat{\Psi}_{2}(X)=\sum_{n\in \Omega _{2}}\mathcal{M}_{n}^{-1/2}a_{n}\ \psi
_{n}(X),\ \ \hat{\Psi}_{4}(X)=\sum_{n\in \Omega _{4}}\mathcal{M}%
_{n}^{-1/2}b_{n}^{\dag }\ \psi _{n}(X),  \label{gen.2}
\end{equation}%
in the ranges $\Omega _{i}$, $i=2,4$

Operators (\ref{op-in}) and (\ref{op-out}) obey the following anticommutation
relations. All operators with different sets of quantum numbers $n$
anticommute. This implies that all operators from different ranges $\Omega
_{i}$ anticommute. Existing inside each range $\Omega _{i}$ are the nonzero anticommutation relations
\begin{align}
& \left[ \ _{+}a_{n},\ _{+}a_{n^{\prime }}^{\dag }\right]
_{+}=\left[ \ ^{-}a_{n},\ ^{-}a_{n^{\prime }}^{\dag }\right] _{+}=\ \left[ \
_{-}a_{n},\ _{-}a_{n^{\prime }}^{\dag }\right] _{+}=\ \left[ \ ^{+}a_{n},\
^{+}a_{n^{\prime }}^{\dag }\right] _{+}=\delta _{nn^{\prime }}, \ n\in \Omega _{1},  \notag \\
& \left[ \ ^{-}a_{n},\ ^{-}a_{n^{\prime }}^{\dag }\right]
_{+}=\ \left[ \ _{-}b_{n},\ _{-}b_{n^{\prime }}^{\dag }\right] _{+}=\ \left[
\ ^{+}a_{n},\ ^{+}a_{n^{\prime }}^{\dag }\right] _{+}=\ \left[ \ _{+}b_{n},\
_{+}b_{n^{\prime }}^{\dag }\right] _{+}=\delta _{nn^{\prime }}, \ n\in \Omega _{3}, \notag \\
&\left[ \ ^{+}b_{n},\ ^{+}b_{n^{\prime }}^{\dag }\right]
_{+}=\ \left[ \ _{-}b_{n},\ _{-}b_{n^{\prime }}^{\dag }\right] _{+}=\ \left[
\ _{+}b_{n},\ _{+}b_{n^{\prime }}^{\dag }\right] _{+}=\ \left[ \ ^{-}b_{n},\
^{-}b_{n^{\prime }}^{\dag }\right] _{+}=\delta _{nn^{\prime }}, \  n\in \Omega _{5}
\label{gen.3}
\end{align}%
in the ranges $\Omega _{i}$, $i=1,3,5$, and%
\begin{align}
&\left[ a_{n},a_{n^{\prime }}^{\dag }\right] _{+}=\delta
_{nn^{\prime }},\ n\in \Omega _{2}, \nonumber \\ 
&\left[ b_{n},b_{n^{\prime }}^{\dag }%
\right] _{+}=\delta _{nn^{\prime }}, \ n\in \Omega _{4}  \label{gen.4}
\end{align}%
in the ranges $\Omega _{i}$, $i=2,4$. Initial and final vacuum vectors are
defined as state vectors annihilated by operators of initial and final
particles:%
\begin{align}
& \ _{+}a_{n}\left\vert 0,\text{\textrm{in}}\right\rangle =\
^{-}a_{n}\left\vert 0,\text{\textrm{in}}\right\rangle =\ _{-}b_{n}\left\vert
0,\text{\textrm{in}}\right\rangle =\ ^{+}b_{n}\left\vert 0,\text{\textrm{in}}%
\right\rangle =0,  \notag \\
& \ _{-}a_{n}\left\vert 0,\text{\textrm{out}}\right\rangle =\
^{+}a_{n}\left\vert 0,\text{\textrm{out}}\right\rangle =\
_{+}b_{n}\left\vert 0,\text{\textrm{out}}\right\rangle =\
^{-}b_{n}\left\vert 0,\text{\textrm{out}}\right\rangle =0,  \label{gen.5}
\end{align}
for quantum numbers from ranges $\Omega_{i}$, $i=1,3,5$, and%
\begin{equation}
n\in\Omega_{2}:\ a_{n}\left\vert 0,\text{\textrm{in}}\right\rangle
=a_{n}\left\vert 0,\text{\textrm{out}}\right\rangle =0,\ n\in\Omega _{4}:\
b_{n}\left\vert 0,\text{\textrm{in}}\right\rangle =\ b_{n}\left\vert 0,\text{%
\textrm{out}}\right\rangle =0,  \label{gen.6}
\end{equation}
in ranges $\Omega_{2}$ and $\Omega_{4}$. Since all operators from different $%
\Omega_{i}$ anticommute, the total initial and final vacua vectors can be
represented as the tensor product
\begin{equation}
\left\vert 0,\text{\textrm{in}}\right\rangle =\otimes\prod_{1,3,5}\left\vert
0,\text{\textrm{in}}\right\rangle ^{(i)}\otimes\left\vert 0\right\rangle
^{(2)}\otimes\left\vert 0\right\rangle ^{(4)},\ \left\vert 0,\text{\textrm{%
out}}\right\rangle =\otimes\prod_{1,3,5}\left\vert 0,\text{\textrm{out}}%
\right\rangle ^{(i)}\otimes\left\vert 0\right\rangle ^{(2)}\otimes\left\vert
0\right\rangle ^{(4)},  \label{gen.7}
\end{equation}
where $\left\vert 0,\text{\textrm{in}}\right\rangle ^{(i)}$ and $\left\vert 0,\text{\textrm{out}}\right\rangle ^{(i)}$  
denote partial \textrm{in} and \textrm{out} vacua in ranges $\Omega_{i}$, $%
i=1,3,5$, and $\left\vert 0\right\rangle ^{(2)}$ and $\left\vert 0\right\rangle
^{(4)}$ partial vacua in ranges $\Omega_{2}$ and $\Omega_{4}$
respectively, 
\begin{equation}
\left\vert 0\right\rangle ^{(2)}=\left\vert 0,\text{\textrm{in}}%
\right\rangle ^{(2)}=\left\vert 0,\text{\textrm{out}}\right\rangle ^{(2)},\
\ \left\vert 0\right\rangle ^{(4)}=\left\vert 0,\text{\textrm{in}}%
\right\rangle ^{(4)}=\left\vert 0,\text{\textrm{out}}\right\rangle ^{(4)}.
\label{gen.7a}
\end{equation}
In addition, inside each range $\Omega_{i}$ the partial vacua can be presented in turn
as the tensor products in quantum modes:%
\begin{equation}
\left\vert 0,\text{\textrm{in}}\right\rangle
^{(i)}=\prod_{n\in\Omega_{i}}\left\vert 0,\text{\textrm{in}}\right\rangle
_{n}^{(i)},\ \left\vert 0,\text{\textrm{out}}\right\rangle
^{(i)}=\prod_{n\in\Omega_{i}}\left\vert 0,\text{\textrm{out}}\right\rangle
_{n}^{(i)},\ \left\vert 0\right\rangle
^{(2,4)}=\prod_{n\in\Omega_{2,4}}\left\vert 0\right\rangle _{n}^{(2,4)}.
\label{gen.8}
\end{equation}
Each of these partial vacuum vectors is destroyed only by annihilation
operators with the corresponding quantum numbers $n$. The \textrm{in} and 
\textrm{out} sets of operators of creation and annihilation of electrons and
positrons as well as \textrm{in} and \textrm{out} vacua are connected via
the special unitary evolution operators $V$ ($VV^{\dag }=I$), $\left\vert 0,%
\text{\textrm{in}}\right\rangle =V\left\vert 0,\text{\textrm{out}}%
\right\rangle ,$%
\begin{equation}
\left\{ a(\text{\textrm{in}}),\ a^{\dag }(\text{\textrm{in}}),\ b(\text{%
\textrm{in}}),\ b^{\dag }(\text{\textrm{in}})\right\} =V\left\{ a(\text{%
\textrm{out}}),\ a^{\dag }(\text{\textrm{out}}),\ b(\text{\textrm{out}}),\
b^{\dag }(\text{\textrm{out}})\right\} V^{\dag }.  \label{gen.9}
\end{equation}%
This in particular implies that%
\begin{equation}
\hat{F}(\text{\textrm{in}})=V\hat{F}(\text{\textrm{out}})V^{\dag },\ \ 
\label{gen.9a}
\end{equation}%
where $\hat{F}($\textrm{in}$)$ is an operator-valued function written in
terms of the \textrm{in} set of the operators of creation and annihilation
operators while $\hat{F}($\textrm{out}$)$ is the same function written in
terms of the \textrm{out} set. The explicit form of the operator $V$ is given
in Ref. \cite{x-case}. The initial partial vacuum states remain vacua \cite%
{x-case} in ranges $\Omega _{1,2,4,5}$ (i.e., the vacuum is stable in these
ranges). In other words,%
\begin{equation}
\left\vert 0,\text{\textrm{in}}\right\rangle ^{(i)}=\left\vert 0,\text{%
\textrm{out}}\right\rangle ^{(i)},\ i=1,2,4,5.  \label{gen.11}
\end{equation}%
We can also define the vacuum-to-vacuum transition amplitude as%
\begin{equation}
c_{\mathrm{v}}=\langle 0,\text{\textrm{out}}\left\vert 0,\text{\textrm{in}}%
\right\rangle =\ ^{(3)}\langle 0,\text{\textrm{out}}\left\vert 0,\text{%
\textrm{in}}\right\rangle ^{(3)}.  \label{gen.12}
\end{equation}%
Taking into account relations (\ref{gen.8}), we can also introduce partial
transition amplitudes for each quantum mode $n$,%
\begin{equation}
c_{\mathrm{v},n}=\ _{n}^{(3)}\langle 0,\text{\textrm{out}}\left\vert 0,\text{%
\textrm{in}}\right\rangle _{n}^{(3)},\ \ c_{\mathrm{v}}=\prod_{n\in \Omega
_{3}}c_{\mathrm{v},n}.  \label{gen.12a}
\end{equation}
The connection between \textrm{in}- and \textrm{out}-operators can be presented 
also via the linear canonical transformation (also called Bogolubov
transformation), which has the following form in different
ranges of quantum numbers $\Omega _{i}$. In the range $\Omega _{1}$ for
electrons the transformation reads
\begin{align}
& \ ^{+}a_{n}=\eta _{\mathrm{R}}g\left( _{+}|^{+}\right) ^{-1}\
_{+}a_{n}+g\left( ^{-}|_{-}\right) ^{-1}g\left( ^{+}|_{-}\right) \ ^{-}a_{n},
\notag \\
& \ _{-}a_{n}=g\left( _{+}|^{+}\right) ^{-1}g\left( _{-}|^{+}\right) \
_{+}a_{n}-\eta _{\mathrm{L}}g\left( ^{-}|_{-}\right) ^{-1}\ ^{-}a_{n}, 
\notag \\
& \ _{+}a_{n}=g\left( _{-}|^{-}\right) ^{-1}g\left( _{+}|^{-}\right) \
_{-}a_{n}+\ \eta _{\mathrm{L}}g\left( ^{+}|_{+}\right) ^{-1}\ ^{+}a_{n}, 
\notag \\
& \ ^{-}a_{n}=-\eta _{\mathrm{R}}g\left( _{-}|^{-}\right) ^{-1}\
_{-}a_{n}+g\left( ^{+}|_{+}\right) ^{-1}g\left( ^{-}|_{+}\right) \ ^{+}a_{n}.
\label{gen.13}
\end{align}%
The parameters $\ \eta _{\mathrm{L(R)}}=$ \textrm{sgn}$\left( p_{0}-U_{\mathrm{L(R)}}\right) $ 
denote the signs of asymptotic particle kinetic energy.
Canonical transformations between the initial and final pairs of creation
operators of positrons in $\Omega _{5}$ can be derived from the 
expression~(\ref{gen.13}) by replacing $^{\pm }a_{n}\rightarrow \ ^{\pm }b_{n}^{\dag }$%
,\ $_{\pm }a_{n}\rightarrow \ _{\pm }b_{n}^{\dag }$, and $\eta _{\mathrm{L}%
}\rightleftarrows \eta _{\mathrm{R}}$. In the Klein zone $\Omega _{3}$, the
canonical transformation takes the form%
\begin{align}
& \ ^{+}a_{n}=-g\left( _{-}|^{+}\right) ^{-1}\ _{-}b_{n}^{\dag }+g\left(
^{-}|_{+}\right) ^{-1}g\left( ^{+}|_{+}\right) \ ^{-}a_{n},  \notag \\
& \ _{+}b_{n}^{\dag }=g\left( _{-}|^{+}\right) g\left( _{+}|^{+}\right)
^{-1}\ _{-}b_{n}^{\dag }+g\left( ^{-}|_{+}\right) ^{-1}\ ^{-}a_{n},  \notag
\\
& \ _{-}b_{n}^{\dag }=g\left( _{+}|^{-}\right) ^{-1}g\left( _{-}|^{-}\right)
\ _{+}b_{n}^{\dag }-g\left( ^{+}|_{-}\right) ^{-1}\ ^{+}a_{n},  \notag \\
& \ ^{-}a_{n}=g\left( _{+}|^{-}\right) ^{-1}\ _{+}b_{n}^{\dag }+g\left(
^{+}|_{-}\right) ^{-1}g\left( ^{-}|_{-}\right) \ ^{+}a_{n}.  \label{gen.14}
\end{align}%
The functions $g$ are mutual decomposition coefficients of Dirac equation
solutions,%
\begin{align}
& \eta _{\mathrm{L}}\ ^{\zeta }\psi _{n}(X)=\ _{+}\psi _{n}(X)g\left(
_{+}|^{\zeta }\right) -\ _{-}\psi _{n}(X)g\left( _{-}|^{\zeta }\right) ,\  
\notag \\
& \eta _{\mathrm{R}}\ _{\zeta }\psi _{n}(X)=\ ^{+}\psi _{n}(X)g\left(
^{+}|_{\zeta }\right) -\ ^{-}\psi _{n}(X)g\left( ^{-}|_{\zeta }\right) ,\ 
\label{gf.decomposition}
\end{align}%
with respect to the inner product on the $x$-constant hyperplane (see Ref. \cite%
{x-case} for details), and have the following properties:%
\begin{align}
& \left( _{\zeta }\psi _{n},^{\zeta ^{\prime }}\psi _{n^{\prime }}\right)
_{x}=\delta _{n,n^{\prime }}g\left( _{\zeta }|^{\zeta ^{\prime }}\right) ,\
\ g\left( _{\zeta }|^{\zeta ^{\prime }}\right) =g\left( ^{\zeta ^{\prime
}}|_{\zeta }\right) ^{\ast },  \notag \\
& \left\vert g\left( _{-}|^{+}\right) \right\vert ^{2}=\left\vert g\left(
_{+}|^{-}\right) \right\vert ^{2},\ \ \left\vert g\left( _{+}|^{+}\right)
\right\vert ^{2}=\left\vert g\left( _{-}|^{-}\right) \right\vert ^{2},\ \ 
\frac{g\left( _{+}|^{-}\right) }{g\left( _{-}|^{-}\right) }=\frac{g\left(
^{+}|_{-}\right) }{g\left( ^{+}|_{+}\right) }.  \label{gf.dec.prop}
\end{align}

\subsection*{Generating functionals for density operators}

We introduce special generating functionals that allow us to obtain the
explicit forms of density operators (matrices) for different initial states
by choosing an appropriate set of sources. Note that the results of this
section are valid for any $x$-electric potential step. As we mentioned in
the preceding section, all the creation and annihilation operators (\ref{op-in})
from different ranges $\Omega_{i}$ anticommute. The density operator $\hat{\rho}$
of the system under consideration is a function of quadratic combinations of these
creation and annihilation operators. This fact allows us to present the density
operator $\hat{\rho}$ as a tensor product of partial density operators $\hat{%
\rho}^{(i)}$ for each range $\Omega_{i}$, 
\begin{equation}
\hat{\rho}=\otimes\prod_{i=1}^{5}\hat{\rho}^{(i)}.  \label{gf.0}
\end{equation}
One can see that due to Eqs. (\ref{gen.3}) and (\ref{gen.4}), the operators $\hat{%
\rho}^{(i)}$ anticommute and can be considered separately. Thus, it is
convenient to introduce the separate partial generating functional for each
range of quantum numbers $\Omega_{i}$. We will refer to each of these
generating functionals as $R^{(i)}(J)$, and $J=\left\{ J_{n}\right\}
_{n\in\Omega_{i}}$ is a complete set of sources in each range which fully
describes (parametrizes) the initial state of the system in each range $%
\Omega_{i}$. The total generating functional can be obtained as a direct
tensor product of functionals $R^{(i)}(J)$,%
\begin{equation}
R(J)=\otimes\prod_{i=1}^{5}R^{(i)}(J).  \label{gf.1}
\end{equation}

\subsubsection*{Generating functionals in $\Omega _{1}$ and $\Omega _{5}$.}

\noindent In $\Omega_{1}$ the generating functional $R^{(1)}(J)$ has the form%
\begin{align}
& R^{(1)}(J)=\prod_{n\in\Omega_{1}}R_{n}^{(1)},\ \ R_{n}^{(1)}=\left[
Z_{n}^{(1)}\right] ^{-1}\ \underline{R}_{n}^{(1)},\ \ \text{\textrm{tr}}%
R_{n}^{(1)}=1,  \notag \\
& \underline{R}_{n}^{(1)}=\mathbf{:}\exp\left[ \ _{+}a_{n}^{\dag}\left(
J_{+,n}^{(1)}-1\right) \ _{+}a_{n}\ +\ ^{-}a_{n}^{\dag}\left(
J_{-,n}^{(1)}-1\right) \ ^{-}a_{n}\ \right] \mathbf{:\ }.  \label{gf.o1}
\end{align}
In $\Omega_{5}$ the generating functional $R^{(5)}(J)$ has the form%
\begin{align}
& R^{(5)}(J)=\prod_{n\in\Omega_{5}}R_{n}^{(5)},\ \ R_{n}^{(5)}=\left[
Z_{n}^{(5)}\right] ^{-1}\ \underline{R}_{n}^{(5)}\mathrm{,\ \ }\text{\textrm{%
tr}}R_{n}^{(5)}=1,  \notag \\
& \underline{R}_{n}^{(5)}=\mathbf{:}\exp\left[ \ ^{+}b_{n}^{\dag}\left(
J_{+,n}^{(5)}-1\right) \ ^{+}b_{n}\ +\ _{-}b_{n}^{\dag}\left(
J_{-,n}^{(5)}-1\right) \ _{-}b_{n}\ \right] \mathbf{:\ }.  \label{gf.o5}
\end{align}
Here $Z_{n}^{(1)}$ and $Z_{n}^{(5)}$ are normalization factors
(statistical sums); colons $:\ldots:$ always denote the normal form with
respect to creation and annihilation operators inside them. Using Eq. (\ref{A.6}), one can calculate each of them as%
\begin{equation}
Z_{n}^{(1,5)}=\left( J_{+,n}^{(1,5)}+1\right) \left(
J_{-,n}^{(1,5)}+1\right) .  \label{gf.Z15}
\end{equation}

\subsubsection*{Generating functionals in $\Omega _{2}$ and $\Omega _{4}$.}

\noindent In these ranges the corresponding generating functionals $%
R^{(2,4)}(J)$ have the following structure:%
\begin{align}
& R^{(2)}(J)=\prod\limits_{n\in\Omega_{2}}R_{n}^{(2)},\ \ R_{n}^{(2)}=\left[
Z_{n}^{(2)}\right] ^{-1}\ \mathbf{:}\exp\left[ a_{n}^{%
\dagger}(J_{n}^{(2)}-1)a_{n}\right] \mathbf{:\ },\   \notag \\
& R^{(4)}(J)=\prod\limits_{n\in\Omega_{4}}R_{n}^{(4)},\ \ R_{n}^{(4)}=\left[
Z_{n}^{(4)}\right] ^{-1}\ \mathbf{:}\exp\left[ b_{n}^{%
\dagger}(J_{n}^{(4)}-1)b_{n}\right] \mathbf{:\ }.\   \label{gf.4}
\end{align}
Here $J_{n}^{(2)}$ and $J_{n}^{(4)}$ are the corresponding sources in $%
\Omega_{2}$ and $\Omega_{4}$, and the corresponding normalization factors are%
\begin{equation}
Z_{n}^{(2,4)}=(J_{n}^{(2,4)}+1).  \label{gf.4a}
\end{equation}
The structure of operators $R^{(2,4)}(J)$ are trivial as there is no particle
production in these ranges and all initial particles are subjected to total
reflection \cite{x-case}. For this reason we often omit the consideration of
ranges $\Omega_{2}$ and $\Omega_{4}$ throughout the article.

\subsubsection*{Generating functional in Klein zone.}

\noindent In the Klein zone, $\Omega _{3}$, the corresponding generating
functional $R^{(3)}(J)$ has the form 
\begin{align}
& R^{(3)}(J)=\prod\limits_{n\in \Omega _{3}}R_{n}^{(3)},\ \ R_{n}^{(3)}= 
\left[ Z_{n}^{(3)}\right] ^{-1}\underline{R}_{n}^{(3)},\mathrm{\ }\text{%
\textrm{tr}}R_{n}^{(3)}=1,  \notag \\
& \underline{R}_{n}^{(3)}=\mathbf{:}\exp \left[ \ ^{-}a_{n}^{\dagger }\left(
J_{+,n}^{(3)}-1\right) \ ^{-}a_{n}+\ _{-}b_{n}^{\dagger }\left(
J_{-,n}^{(3)}-1\right) \ _{-}b_{n}\right] \mathbf{:}\ ,  \label{gf.o3}
\end{align}%
where the normalization factor $Z_{n}^{(3)}$ has the form%
\begin{equation}
Z_{n}^{(3)}=\left( J_{+,n}^{(3)}+1\right) \left( J_{-,n}^{(3)}+1\right) .
\label{gf.6}
\end{equation}

\subsection*{Normal form of generating functional}

The problem of calculating the mean value $F($\textrm{out}$)$ of an operator 
$\hat{F}($\textrm{out}$)$ at the final state of the system is related to the
problem of calculating the quantity \textrm{tr}$\left[ \hat{F}(\mathrm{out})%
\hat{\rho}\right] $, which is 
\begin{align}
& \mathrm{tr}\left[ \hat{F}(\mathrm{out})\hat{\rho}\right] =\sum
_{M,N=0}^{\infty}\sum_{M!N!}\left\langle \Psi_{\left\{ m\right\}
_{M};\left\{ n\right\} _{N}}(\text{\textrm{out}})\right\vert \hat {F}(%
\mathrm{out})\hat{\rho}\left\vert \Psi_{\left\{ m\right\} _{M};\left\{
n\right\} _{N}}(\text{\textrm{out}})\right\rangle ,  \notag \\
& \left\vert \Psi_{\left\{ m\right\} _{M};\left\{ n\right\} _{N}}(\text{%
\textrm{out}})\right\rangle =\ a_{m_{1}}^{\dag}\ldots a_{m_{M}}^{\dag
}b_{n_{1}}^{\dag}\ldots b_{n_{N}}^{\dag}\left\vert 0,\text{\textrm{out}}%
\right\rangle .  \label{gf.6a}
\end{align}
For this reason, it is convenient to have the expression for generating
functionals $R^{(i)}(J)$ (and subsequently for density operators $\hat{%
\rho }$) in terms of the \textrm{out} set of creation and annihilation operators.
According to (\ref{gen.9a}),%
\begin{equation*}
\underline{R}(J)=VU(J)V^{\dag},
\end{equation*}
where $U(J)$ are operators $\underline{R}(J)$ with creation and annihilation
operators from the \textrm{in} set replaced by the corresponding operators from 
\textrm{out} set. Taking into account Eq.~(\ref{gf.1}) and the fact that
evolution operator $V$ can also be factorized as \cite{x-case}%
\begin{equation*}
V=\otimes\prod_{i=1}^{5}V^{(i)},
\end{equation*}
we can write that%
\begin{equation}
\underline{R}^{(i)}(J)=V^{(i)}U^{(i)}(J)V^{(i)\dag},  \label{gf.6b}
\end{equation}
where $U^{(i)}(J)$ are operators $\underline{R}^{(i)}(J)$ with
creation-annihilation operators from \textrm{in}-set replaced by
corresponding operators from \textrm{out}-set for each range $\Omega_{i}$.
Utilizing the explicit forms of operators $V^{(i)}$ , $i=1,3,5$, found in
Ref. \cite{x-case}, we can construct the expression for generating
functionals $R^{(i)}(J)$ in terms of the \textrm{out} set. It should be noted
that the unitary evolution operators $V^{(i)}$ have the same functional form
in terms of the \textrm{in} and \textrm{out} sets of operators of particle
creation and annihilation due to the properties (\ref{gen.9}) and (\ref{gen.9a}).

\subsubsection*{Ranges $\Omega _{1}$ and $\Omega _{5}$}

\noindent In $\Omega_{1}$ the partial evolution operator $V^{(1)}=\prod
_{n\in\Omega_{1}}V_{n}^{(1)}$ has the form 
\begin{align}
& V_{n}^{(1)}=\exp\left[ \ ^{+}a_{n}^{\dag}\ S_{4}\ _{-}a_{n}\right] \ \exp%
\left[ \ _{-}a_{n}^{\dag}\ S_{3}\ _{-}a_{n}\right] \exp\left[ \
^{+}a_{n}^{\dag}\ S_{2}\ ^{+}a_{n}\right] \exp\left[ \ _{-}a_{n}^{\dag }\
S_{1}\ ^{+}a_{n}\right] ,  \notag \\
& S_{4}=g\left( _{-}|^{+}\right) ^{-1},\ S_{3}=\ln\left[ g\left(
^{-}|_{-}\right) ^{-1}g\left( ^{-}|_{+}\right) \right] ,\ S_{2}=\ln\left[
g\left( ^{+}|_{+}\right) g\left( ^{-}|_{+}\right) ^{-1}\right] ,\
S_{1}=-g\left( ^{-}|_{+}\right) ^{-1}.  \label{gf.2}
\end{align}
In $\Omega_{5}$ the operator $V^{(5)}=\prod\limits_{n\in%
\Omega_{5}}V_{n}^{(5)}$ is%
\begin{align}
& V_{n}^{(5)}=\exp\left[ \ _{+}b_{n}^{\dagger}\ S_{4}^{\prime}\ ^{-}b_{n}%
\right] \exp\left[ \ ^{-}b_{n}^{\dagger}\ S_{3}^{\prime}\ ^{-}b_{n}\right]
\exp\left[ \ _{+}b_{n}^{\dagger}\ S_{2}^{\prime}\ _{+}b_{n}\right] \exp\left[
\ ^{-}b_{n}^{\dagger}\ S_{1}^{\prime}\ _{+}b_{n}\right] ,  \notag \\
& S_{4}^{\prime}=-g(_{+}|^{-})^{-1},\ S_{3}^{\prime}=\ln\left[
g(^{-}|_{-})^{-1}g(^{+}|_{-})\right] ,\ S_{2}^{\prime}=\ln\left[
g(^{+}|_{+})g(^{+}|_{-})^{-1}\right] ,\ S_{1}^{\prime}=g(^{+}|_{-})^{-1}.
\label{gf.3}
\end{align}
Then one can write that%
\begin{equation}
\underline{R}_{n}^{(1)}=V_{n}^{(1)}U_{n}^{(1)}V_{n}^{(1)\dagger},
\label{r1out}
\end{equation}
where $U_{n}^{(1)}$ is the operator that can be obtained from $\underline{R}%
_{n}^{(1)}$ by simultaneous replacements $\ _{+}a_{n}\rightarrow\ ^{+}a_{n}$
and $\ ^{-}a_{n}\rightarrow\ _{-}a_{n}$. Similarly, we have that 
\begin{equation}
\underline{R}_{n}^{(5)}=V_{n}^{(5)}U_{n}^{(5)}V_{n}^{(5)\dagger},
\label{r5out}
\end{equation}
where $U_{n}^{(5)}$ is the operator that can be obtained from $\underline{R}%
_{n}^{(5)}$ by simultaneous replacements $\ _{-}b_{n}\rightarrow\ ^{-}b_{n}$
and $\ ^{+}b_{n}\rightarrow\ _{+}b_{n}$. Let us calculate the normal form of
the operator $\underline{R}_{n}^{(1)}$. This can be done using the relation (\ref{A.3}). 
Utilizing the anticommutation relations for the
creation and annihilation operators (\ref{gen.3}), one can find that%
\begin{align}
& \exp\left[ \ ^{+}a_{n}^{\dag}\ S_{2}\ ^{+}a_{n}\right] \exp\left[ \
_{-}a_{n}^{\dag}\ S_{1}\ ^{+}a_{n}\right] =\exp\left[ \ _{-}a_{n}^{\dag }\
S_{1}e^{-S_{2}}\ ^{+}a_{n}\right] \exp\left[ \ ^{+}a_{n}^{\dag}\ S_{2}\
^{+}a_{n}\right] ,  \notag \\
& \exp\left[ \ _{-}a_{n}^{\dag}\ S_{3}\ _{-}a_{n}\right] \exp\left[ \
_{-}a_{n}^{\dag}\ S_{1}e^{-S_{2}}\ ^{+}a_{n}\right] =\exp\left[ \
_{-}a_{n}^{\dag}e^{S_{3}}\ S_{1}e^{-S_{2}}\ ^{+}a_{n}\right] \exp\left[ \
_{-}a_{n}^{\dag}\ S_{3}\ _{-}a_{n}\right] .  \label{gf.3b}
\end{align}
Then partial operators $V_{n}^{(1)}$ with the help of the relation (\ref{A.2})
can be presented as%
\begin{align}
& V_{n}^{(1)}=Y_{n}\tilde{V}_{n},\ Y_{n}=\exp\left[ \ ^{+}a_{n}^{\dag }\
S_{4}\ _{-}a_{n}\right] \exp\left[ \ _{-}a_{n}^{\dag}e^{S_{3}}\
S_{1}e^{-S_{2}}\ ^{+}a_{n}\right] ,  \notag \\
& \tilde{V}_{n}=\mathbf{:}\exp\left[ \ ^{+}a_{n}^{\dag}\ \left(
e^{S_{2}}-1\right) \ ^{+}a_{n}+\ _{-}a_{n}^{\dag}\ \left( e^{S_{3}}-1\right)
\ _{-}a_{n}\right] \mathbf{:}\ .  \label{gf.3c}
\end{align}
Using the relation (\ref{A.4}), one can present the product $\tilde{V}%
_{n}U_{n}^{(1)}\tilde{V}_{n}^{\dag}$ as follows:%
\begin{align}
& \tilde{V}_{n}U_{n}^{(1)}\tilde{V}_{n}^{\dag}=\ \mathbf{:}\exp\left[ \
^{+}a_{n}^{\dag}\ A_{++}\ ^{+}a_{n}+\ _{-}a_{n}^{\dag}\ A_{--}\ _{-}a_{n}%
\right] \mathbf{:}\ ,  \notag \\
& A_{++}=J_{n,+}^{(1)}\left\vert e^{S_{2}}\right\vert ^{2}-1,\ \
A_{--}=J_{n,-}^{(1)}\left\vert e^{S_{3}}\right\vert ^{2}-1.  \label{gf.3d}
\end{align}
On the other hand, with the help of relation (\ref{A.5}), the operator $%
Y_{n} $ can be presented as%
\begin{equation}
Y_{n}=\ \mathbf{:}\exp\left[ \ ^{+}a_{n}^{\dag}\ S_{4}\ _{-}a_{n}+\
_{-}a_{n}^{\dag}\ \tilde{S}_{1}\ ^{+}a_{n}+\ ^{+}a_{n}^{\dag}\ S_{4}\tilde{S}%
_{1}\ ^{+}a_{n}\right] \mathbf{:},\ \tilde{S}_{1}=e^{S_{3}}\ S_{1}e^{-S_{2}},
\label{gf.3e}
\end{equation}
Then one can calculate $Y_{n}\tilde{V}_{n}U_{n}^{(1)}\tilde{V}_{n}^{\dag}$
to be%
\begin{align}
& Y_{n}\tilde{V}_{n}U_{n}^{(1)}\tilde{V}_{n}^{\dag}=\mathbf{:}\exp\left[ \
^{+}a_{n}^{\dag}\ B_{+-}\ _{-}a_{n}+\ _{-}a_{n}^{\dag}\ B_{-+}\ ^{+}a_{n}+\
^{+}a_{n}^{\dag}\ B_{++}\ ^{+}a_{n}\ +\ _{-}a_{n}^{\dag}\ B_{--}\ _{-}a_{n}%
\right] \mathbf{:},  \notag \\
& B_{++}=A_{++}+\left( 1+A_{++}\right) S_{4}\tilde{S}_{1},\ \
B_{--}=A_{--},\ B_{+-}=S_{4}\left( 1+A_{--}\right) ,\ \ B_{-+}=\left(
1+A_{++}\right) \tilde{S}_{1}.  \label{gf.3f}
\end{align}
Finally, we can attach the last remaining operator $Y_{n}^{\dag}$ from the
right side, to obtain 
\begin{align}
& \underline{R}_{n}^{(1)}=\mathbf{:}\exp\left[ \ ^{+}a_{n}^{\dag}\ C_{+-}\
_{-}a_{n}+\ _{-}a_{n}^{\dag}\ C_{-+}\ ^{+}a_{n}+\ ^{+}a_{n}^{\dag }\ C_{++}\
^{+}a_{n}\ +\ _{-}a_{n}^{\dag}\ C_{--}\ _{-}a_{n}\right] \mathbf{:},  \notag
\\
& C_{++}=S_{4}^{\ast}\tilde{S}_{1}^{\ast}+B_{+-}S_{4}^{\ast}+B_{++}\left(
1+S_{4}^{\ast}\tilde{S}_{1}^{\ast}\right) ,\ \ C_{--}=B_{--}+B_{-+}\tilde {S}%
_{1}^{\ast},  \notag \\
& C_{-+}=S_{4}^{\ast}+B_{-+}+B_{-+}S_{4}^{\ast}\tilde{S}_{1}^{%
\ast}+B_{--}S_{4}^{\ast},\ \ C_{+-}=\tilde{S}_{1}^{\ast}+B_{+-}+B_{++}\tilde{%
S}_{1}^{\ast}.  \label{gf.3g}
\end{align}
Substituting $B_{\zeta\zeta^{\prime}}$ and $A_{\pm\pm}$ into the 
expression (\ref{gf.3g}), we find the explicit form of the operator $R^{(1)}$ in terms of 
\textrm{out} operators to be%
\begin{align}
& C_{++}=-1+J_{n,-}^{(1)}\left\vert S_{4}\right\vert ^{2}\left\vert
e^{S_{3}}\right\vert ^{2}+J_{n,+}^{(1)}\left( 1+S_{4}\tilde{S}_{1}\right)
\left( 1+S_{4}^{\ast}\tilde{S}_{1}^{\ast}\right) \left\vert
e^{S_{2}}\right\vert ^{2},  \notag \\
& C_{--}=-1+J_{n,-}^{(1)}\left\vert e^{S_{3}}\right\vert
^{2}+J_{n,+}^{(1)}\left\vert \tilde{S}_{1}\right\vert ^{2}\left\vert
e^{S_{2}}\right\vert ^{2},\ \   \notag \\
& C_{-+}=J_{n,+}^{(1)}\tilde{S}_{1}\left( 1+S_{4}^{\ast}\tilde{S}_{1}^{\ast
}\right) \left\vert e^{S_{2}}\right\vert ^{2}+J_{n,-}^{(1)}S_{4}^{\ast
}\left\vert e^{S_{3}}\right\vert ^{2},  \notag \\
& C_{+-}=J_{n,-}^{(1)}S_{4}\left\vert e^{S_{3}}\right\vert ^{2}+J_{n,+}^{(1)}%
\tilde{S}_{1}^{\ast}\left( 1+S_{4}\tilde{S}_{1}\right) \left\vert
e^{S_{2}}\right\vert ^{2}.\ \   \label{gf.3h}
\end{align}
In ranges $\Omega_{1}$ and $\Omega_{5}$ the matrices $g$ are connected \cite%
{x-case} to the relative amplitudes of an electron and positron reflection $%
R_{\pm}$ and transmission $T_{\pm}$ as
\begin{align}
& R_{+,n}=g\left( _{+}|^{+}\right) ^{-1}g\left( _{-}|^{+}\right) ,\ \
T_{+,n}=\eta_{\mathrm{L}}g\left( _{+}|^{+}\right) ^{-1},  \notag \\
& R_{-,n}=g\left( ^{-}|_{-}\right) ^{-1}g\left( ^{+}|_{-}\right) ,\ \
T_{-,n}=-\eta_{\mathrm{R}}g\left( ^{-}|_{-}\right) ^{-1}.  \label{RT}
\end{align}
One can use these definitions and the properties of matrices $g$ given by (\ref%
{gf.dec.prop}) to present the coefficients from Eq. (\ref{gf.3h}) as%
\begin{align}
& 1+S_{4}\tilde{S}_{1}=1-\left\vert g\left( ^{-}|_{-}\right) \right\vert
^{-2}\frac{\left\vert g\left( ^{-}|_{+}\right) \right\vert ^{2}}{\left\vert
g\left( ^{+}|_{-}\right) \right\vert ^{2}}=1-\left\vert g\left(
^{-}|_{-}\right) \right\vert ^{-2}=\left\vert R_{+,n}\right\vert ^{2}, 
\notag \\
& \left\vert S_{4}\right\vert ^{2}=\left\vert g\left( ^{-}|_{+}\right)
\right\vert ^{-2}=\left\vert T_{+,n}\right\vert ^{2}/\left\vert
R_{+,n}\right\vert ^{2},  \notag \\
& \left\vert e^{S_{2}}\right\vert ^{2}=\left\vert g\left( ^{+}|_{+}\right)
\right\vert ^{2}\left\vert g\left( ^{-}|_{+}\right) \right\vert
^{2}=\left\vert R_{+,n}\right\vert ^{-2},  \notag \\
& \left\vert e^{S_{3}}\right\vert ^{2}=\left\vert g\left( ^{-}|_{-}\right)
\right\vert ^{-2}\left\vert g\left( ^{+}|_{-}\right) \right\vert
^{2}=\left\vert R_{+,n}\right\vert ^{2}.  \label{StoRT}
\end{align}
Thus coefficients $C$ in Eq. (\ref{gf.3h}) take the compact form
\begin{align}
& C_{++}=-1+J_{n,-}^{(1)}\left\vert T_{n}\right\vert
^{2}+J_{n,+}^{(1)}\left\vert R_{n}\right\vert ^{2},\
C_{--}=-1+J_{n,-}^{(1)}\left\vert R_{n}\right\vert
^{2}+J_{n,+}^{(1)}\left\vert T_{n}\right\vert ^{2},  \notag \\
& C_{-+}=J_{n,-}^{(1)}g\left( ^{+}|_{-}\right) ^{-1}\left\vert
R_{n}\right\vert ^{2}-J_{n,+}^{(1)}g\left( _{-}|^{+}\right) \left\vert
T_{n}\right\vert ^{2},  \notag \\
& C_{+-}=J_{n,-}^{(1)}g\left( _{-}|^{+}\right) ^{-1}\left\vert
R_{n}\right\vert ^{2}-J_{n,+}^{(1)}g\left( ^{+}|_{-}\right) \left\vert
T_{n}\right\vert ^{2},  \label{CtoRT}
\end{align}
where we introduced the notation
\begin{equation}
\left\vert T_{n}\right\vert ^{2}=\left\vert T_{-,n}\right\vert
^{2}=\left\vert T_{+,n}\right\vert ^{2},\ \left\vert R_{n}\right\vert
^{2}=\left\vert R_{-,n}\right\vert ^{2}=\left\vert R_{+,n}\right\vert ^{2},
\label{RTsqr}
\end{equation}
where $\left\vert T_{n}\right\vert ^{2}$ and $\left\vert R_{n}\right\vert
^{2}$ are the absolute probability of electron transmission and the absolute
probability of electron reflection, respectively, so $\left\vert
T_{n}\right\vert ^{2}+\left\vert R_{n}\right\vert ^{2}=1$. The normal form
of the operator $\underline{R}_{n}^{(5)}$ can be constructed in the same
manner, and has the form%
\begin{align}
& \underline{R}_{n}^{(5)}=:\exp\left[ \ _{+}b_{n}^{\dag}\ D_{++}\ _{+}b_{n}\
+\ _{+}b_{n}^{\dag}\ D_{+-}\ ^{-}b_{n}+\ ^{-}b_{n}^{\dag}\ D_{-+}\
_{+}b_{n}+\ ^{-}b_{n}^{\dag}\ D_{--}\ ^{-}b_{n}\right] :\ ,  \notag \\
& D_{++}=-1+J_{n,-}^{(1)}\left\vert T_{n}\right\vert
^{2}+J_{n,+}^{(1)}\left\vert R_{n}\right\vert ^{2},\
D_{-+}=-J_{n,-}^{(1)}g\left( ^{-}|_{+}\right) ^{-1}\left\vert
R_{n}\right\vert ^{2}+J_{n,+}^{(1)}g\left( _{+}|^{-}\right) \left\vert
T_{n}\right\vert ^{2},  \notag \\
& D_{--}=-1+J_{n,-}^{(1)}\left\vert R_{n}\right\vert
^{2}+J_{n,+}^{(1)}\left\vert T_{n}\right\vert ^{2},\
D_{+-}=-J_{n,-}^{(1)}g\left( _{+}|^{-}\right) ^{-1}\left\vert
R_{n}\right\vert ^{2}+J_{n,+}^{(1)}g\left( ^{-}|_{+}\right) \left\vert
T_{n}\right\vert ^{2}.  \label{gf.3j}
\end{align}

\subsubsection*{Range $\Omega _{3}$}

\noindent The unitary evolution operator $V^{(3)}$ has the form 
\begin{align}
& V^{(3)}=\prod_{n\in \Omega _{3}}V_{n}^{(3)},\
V_{n}^{(3)}=w_{n}(-|-)^{-1}v_{4}^{(3)}v_{3}^{(3)}v_{2}^{(3)}v_{1}^{(3)}, 
\notag \\
& v_{4}^{(3)}=\exp \left\{ -\ ^{+}a_{n}^{\dag }\ w_{n}\left( +-|0\right) \
_{+}b_{n}^{\dag }\right\} ,\ v_{3}^{(3)}=\exp \left\{ \ _{+}b_{n}^{\dag }\
\ln w_{n}(-|-)\ _{+}b_{n}\right\} ,  \notag \\
& v_{2}^{(3)}=\exp \left\{ \ ^{+}a_{n}\ \ln w_{n}(+|+)\ ^{+}a_{n}\right\} ,\
v_{1}^{(3)}=\exp \left\{ -\ _{+}b_{n}\ w_{n}\left( 0|-+\right) \
^{+}a_{n}\right\} ,  \label{V3}
\end{align}%
where $w_{n}(\zeta |\zeta ^{\prime })$ are elementary amplitudes of
scattering and pair creation processes, defined as 
\begin{align}
& w_{n}(-|-)=c_{\mathrm{v}}^{-1}\left\langle 0,\text{\textrm{out}}%
\right\vert \ _{+}b_{n}\ _{-}b_{n}^{\dag }\left\vert 0,\text{\textrm{in}}%
\right\rangle ,\ \ w_{n}(+|+)=c_{\mathrm{v}}^{-1}\left\langle 0,\text{%
\textrm{out}}\right\vert \ ^{+}a_{n}\ ^{-}a_{n}^{\dag }\left\vert 0,\text{%
\textrm{in}}\right\rangle ,\   \notag \\
& w_{n}\left( +-|0\right) =c_{\mathrm{v}}^{-1}\left\langle 0,\text{\textrm{%
out}}\right\vert \ ^{+}a_{n}\ _{+}b_{n}\left\vert 0,\text{\textrm{in}}%
\right\rangle ,\ \ w_{n}\left( 0|-+\right) =c_{\mathrm{v}}^{-1}\left\langle
0,\text{\textrm{out}}\right\vert \ _{-}b_{n}^{\dag }\ ^{-}a_{n}^{\dag
}\left\vert 0,\text{\textrm{in}}\right\rangle .  \label{omega}
\end{align}%
All these amplitudes are diagonal in quantum numbers due to Eq. (\ref{gen.3}%
) and can be expressed in terms of the coefficients $g\left( ^{\zeta
}|_{\zeta ^{\prime }}\right) $ as follows:%
\begin{align}
& w_{n}(-|-)=g\left( ^{-}|_{+}\right) g\left( ^{-}|_{-}\right) ^{-1}=g\left(
_{-}|^{+}\right) g\left( _{+}|^{+}\right) ^{-1},  \notag \\
& w_{n}(+|+)=g\left( ^{+}|_{-}\right) g\left( ^{-}|_{-}\right) ^{-1}=g\left(
_{+}|^{-}\right) g\left( _{+}|^{+}\right) ^{-1},  \notag \\
& w_{n}\left( +-|0\right) =g\left( _{+}|^{+}\right) ^{-1},\ w_{n}\left(
0|-+\right) =-g\left( ^{-}|_{-}\right) ^{-1}.  \label{omega.g}
\end{align}%
The relative amplitude of pair creation, $w_{n}\left( +-|0\right) ,$ is also
connected with the differential number of pairs created from vacuum, 
\begin{equation}
N_{n}^{\text{$\mathrm{cr}$}}=\left\vert g\left( _{-}|^{+}\right) \right\vert
^{-2}=\frac{\left\vert w_{n}(+-|0)\right\vert ^{2}}{1+\left\vert
w_{n}(+-|0)\right\vert ^{2}},\ \left\vert w_{n}(+-|0)\right\vert ^{2}=\frac{%
N_{n}^{\text{$\mathrm{cr}$}}}{1-N_{n}^{\text{$\mathrm{cr}$}}},\ \left\vert
c_{\mathrm{v},n}\right\vert ^{2}=1-N_{n}^{\text{$\mathrm{cr}$}}.
\label{w.g.n}
\end{equation}
We note that the structure of the operator $V^{(3)}$ can be formally identified with
the structure of the unitary evolution operator $V$ for QED with
time-dependent uniform electric potential steps \cite{density} with the formal
replacements $a_{n}\rightarrow \ ^{+}a_{n}^{\dag }$ and $b_{n}\rightarrow \
_{+}b_{n}^{\dag }$. Thus, the normal form of the operator $\underline{R}%
_{n}^{(3)}$ can be obtained in the exact same way as in Ref. \cite{density} and has the form%
\begin{align}
& \underline{R}_{n}^{(3)}=\left\vert w_{n}(-|-)\right\vert ^{-2}\left(
1+AB\right) :\exp \left[ -\ ^{+}a_{n}^{\dag }\left( 1-D_{+}\right) \
^{+}a_{n}\right.  \notag \\
& \left. -\ _{+}b_{n}^{\dag }\left( 1-D_{-}\right) \ _{+}b_{n}-\
^{+}a_{n}^{\dag }C^{\dag }\ _{+}b_{n}^{\dag }-\ _{+}b_{n}C\ ^{+}a_{n}\right]
:\ ,  \notag \\
& D_{+}=\left\vert w_{n}(+|+)\right\vert ^{2}\left( 1+AB\right)
^{-1}J_{+,n}^{(3)},\ \ B=w_{n}\left( 0|-+\right) ,  \notag \\
& D_{-}=\left\vert w_{n}\left( -|-\right) \right\vert
^{2}J_{-,n}^{(3)}\left( 1+AB\right) ^{-1},\ \ A=J_{+,n}^{(3)}B^{\ast
}J_{-,n}^{(3)},  \notag \\
& C=w_{n}\left( -|-\right) ^{\ast }A^{\ast }\left( 1+AB\right)
^{-1}w_{n}\left( +|+\right) ^{\ast }+w_{n}\left( +-|0\right) ^{\ast }.
\label{gf.7}
\end{align}

\section*{Appendix B}

\appendix\setcounter{equation}{0}\renewcommand{\theequation}{B\arabic{equation}} 

In this appendix, we consider the case of pure initial states other than
vacuum, namely, initial states with a definite number of particles.

\subsection*{Generating functionals}

The generating functionals $R^{(i)}$ also allow us to construct the partial
density operators $\hat{\rho}_{\left\{ m\right\} _{M};\left\{ n\right\}
_{N}}^{(i)}$ of the system which is initially found in a pure
state with a definite number of particles with fixed sets of quantum numbers 
$\left\{ m\right\} _{M}=\left\{ m_{1},m_{2},\ldots ,m_{M}\right\} $ and $%
\left\{ n\right\} _{N}=\left\{ n_{1},n_{2},\ldots ,n_{N}\right\} $ as
follows. In ranges $\Omega _{1}$, $\Omega _{3}$, and $\Omega
_{5}$, 
\begin{align}
& \hat{\rho}_{\left\{ m\right\} _{M};\left\{ n\right\} _{N}}^{(i)}=\left. 
\frac{\partial ^{M+N}\underline{R}^{(i)}(J)}{\partial \left(
J_{+,m_{1}}^{(i)}\ldots J_{+,m_{M}}^{(i)}\ J_{-,n_{1}}^{(i)}\ldots
J_{-,n_{N}}^{(i)}\right) }\right\vert _{J=0}  \notag \\
& =\left\vert \Psi _{\left\{ m\right\} _{M};\left\{ n\right\} _{N}}(\text{%
\textrm{in}})\right\rangle ^{(i)}\ ^{(i)}\left\langle \Psi _{\left\{
m\right\} _{M};\left\{ n\right\} _{N}}(\text{\textrm{in}})\right\vert ,\ \
m,n\in \Omega _{i},  \label{nm.1}
\end{align}%
where the states $\left\vert \Psi _{\left\{ m\right\} _{M};\left\{ n\right\}
_{N}}(\text{\textrm{in}})\right\rangle ^{(i)}$ are defined as 
\begin{align}
& \left\vert \Psi _{\left\{ m\right\} _{M};\left\{ n\right\} _{N}}(\text{%
\textrm{in}})\right\rangle ^{(1)}=_{+}a_{m_{1}}^{\dag }\ldots \
_{+}a_{m_{M}}^{\dag }\ ^{-}a_{n_{1}}^{\dag }\ldots \ ^{-}a_{n_{N}}^{\dag
}\left\vert 0,\text{\textrm{in}}\right\rangle ^{(1)},\ \   \notag \\
& \left\vert \Psi _{\left\{ m\right\} _{M};\left\{ n\right\} _{N}}(\text{%
\textrm{in}})\right\rangle ^{(3)}=\ ^{-}a_{m_{1}}^{\dagger }\ldots \
^{-}a_{m_{M}}^{\dagger }\ _{-}b_{n_{1}}^{\dagger }\ldots \
_{-}b_{n_{N}}^{\dagger }\left\vert 0,\text{\textrm{in}}\right\rangle
^{(3)},\ \   \notag \\
& \left\vert \Psi _{\left\{ m\right\} _{M};\left\{ n\right\} _{N}}(\text{%
\textrm{in}})\right\rangle ^{(5)}=\ ^{+}b_{m_{1}}^{\dag }\ldots \ \
^{+}b_{m_{M}}^{\dag }\ _{-}b_{n_{1}}^{\dagger }\ldots \
_{-}b_{n_{N}}^{\dagger }\left\vert 0,\text{\textrm{in}}\right\rangle
^{(5)}.\ \   \label{nm.2}
\end{align}%
In the ranges $\Omega _{2}$ and $\Omega _{4}$,
\begin{equation}
\hat{\rho}_{\left\{ n\right\} _{N}}^{(i)}=\left. \frac{\partial ^{N}%
\underline{R}^{(i)}(J)}{\partial \left( J_{n_{1}}^{(i)}\ldots
J_{n_{N}}^{(i)}\right) }\right\vert _{J=0}=\left\vert \Psi _{\left\{
n\right\} _{N}}(\text{\textrm{in}})\right\rangle ^{(i)}\ ^{(i)}\left\langle
\Psi _{\left\{ n\right\} _{N}}(\text{\textrm{in}})\right\vert ,\ \ n\in
\Omega _{i},  \label{nm.3}
\end{equation}%
with the states $\left\vert \Psi _{\left\{ n\right\} _{N}}(\text{\textrm{in}}%
)\right\rangle ^{(i)}$ having the form%
\begin{equation}
\left\vert \Psi _{\left\{ n\right\} _{N}}(\text{\textrm{in}})\right\rangle
^{(2)}=\ a_{n_{1}}^{\dag }\ldots \ a_{n_{N}}^{\dag }\left\vert
0\right\rangle ^{(2)},\ \ \left\vert \Psi _{\left\{ n\right\} _{N}}(\text{%
\textrm{in}})\right\rangle ^{(4)}=\ b_{n_{1}}^{\dag }\ldots \
b_{n_{N}}^{\dag }\left\vert 0\right\rangle ^{(4)}.  \label{nm.4}
\end{equation}

\subsection*{Reduced density operators and entropy production}

First, we consider the reduction due to the measurement of the number of final
particles. It should be stressed that the fact that quantum modes evolve
separately substantially simplifies the technical side of the consideration.
Suppose that the initial particles are present in only one quantum mode $%
m\in \Omega _{i}$. In this case, the partial density operator $\hat{\rho}%
^{(i)}$ for the range $\Omega _{i}$ can be presented as 
\begin{equation}
\hat{\rho}^{(i)}=\hat{\rho}_{m}^{(i)}\otimes \prod_{n\neq m\in \Omega _{i}}%
\hat{\rho}_{v,n}^{(i)},  \label{aaa}
\end{equation}%
where $\hat{\rho}_{m}^{(i)}$ is the partial density operator for the quantum
mode $m$ corresponding to the initial state with a definite number of
particles in question. Due to the structure of the operator $\hat{\rho}^{(i)}$
given by Eq. (\ref{aaa}), the operator $\hat{\rho}_{N}^{(i)}$ takes the form 
\begin{equation}
\hat{\rho}_{N}^{(i)}=\hat{\rho}_{N,m}^{(i)}\otimes \prod_{n\neq m\in \Omega
_{i}}\hat{\rho}_{N,n}^{(i)}.  \label{bbb}
\end{equation}%
One can see that the procedure of reduction for the case under consideration
differs from the deformation of the vacuum initial state only in the quantum
mode $m$, where initial particles are present. One can also see that
it is not difficult to generalize the consideration for the case when
initial particles are present in more than one quantum mode. Let us first
consider the deformations for the range $\Omega _{1}$. It is easy to verify
that the only nonzero weights $W_{s,n}^{(1)}$, $n\neq m$, in Eq. (\ref{bbb}) 
are those where $\left\vert s_{i},\text{\textrm{out}}\right\rangle
_{n}^{(1)}=\left\vert 0,\text{\textrm{out}}\right\rangle _{n}^{(1)}$, i.e.,
partial density operators for vacuum quantum modes do not change due to
measurement of the number of particles; it is possible to write that%
\begin{equation}
\hat{\rho}_{N}^{(1)}=\hat{\rho}_{N,m}^{(1)}\otimes \prod_{n\neq m\in \Omega
_{i}}\hat{\rho}_{v,n}^{(1)}.  \label{ccc}
\end{equation}%
Then, all that is left is to deal with is the quantum mode $m$ where initial
particles are present. Constructing the pure states with \textrm{(a)} a
single right initial electron$\ ^{-}a_{m}^{\dag }$ in mode $m$, \textrm{(b)}
a single left initial electron$\ _{+}a_{m}^{\dag }$, and \textrm{(c)} with both
left and right electrons in the initial state, it is easy to obtain the
measurement-reduced density operators, 
\begin{align}
& \hat{\rho}_{N,m}^{(1)}=\left\vert T_{-,m}\right\vert ^{2}\ _{-}a_{m}^{\dag
}\ \hat{\rho}_{v,m}^{(1)}\ _{-}a_{m}+\left\vert R_{-,m}\right\vert ^{2}\
^{+}a_{m}^{\dag }\ \hat{\rho}_{v,m}^{(1)}\ ^{+}a_{m}\ \ \text{\textrm{for (a)%
}},  \notag \\
& \hat{\rho}_{N,m}^{(1)}=\left\vert R_{+,m}\right\vert ^{2}\ _{-}a_{m}^{\dag
}\ \hat{\rho}_{v,m}^{(1)}\ _{-}a_{m}+\left\vert T_{+,m}\right\vert ^{2}\
^{+}a_{m}^{\dag }\ \hat{\rho}_{v,m}^{(1)}\ ^{+}a_{m}\mathrm{\ \ }\text{%
\textrm{for (b)}},  \notag \\
& \hat{\rho}_{N,m}^{(1)}=\left[ \left\vert R_{m}\right\vert ^{2}+\left\vert
T_{m}\right\vert ^{2}\right] ^{2}\ ^{+}a_{m}^{\dag }\ _{-}a_{m}^{\dag }\ 
\hat{\rho}_{v,m}^{(1)}\ _{-}a_{m}\ ^{+}a_{m}  \notag \\
& \ =\ ^{+}a_{m}^{\dag }\ _{-}a_{m}^{\dag }\ \hat{\rho}_{v,m}^{(1)}\
_{-}a_{m}\ ^{+}a_{m}\ \ \text{\textrm{for (c)}}.  \label{rd.13}
\end{align}%
From the expression (\ref{rd.13}) one can see that for the case of a single initial
electron we have two terms: The first term corresponds to the reflection of the
initial electron, while the second term corresponds to the transition of the
electron through the barrier. When there are two electrons in the initial
state, we see that they can be either simultaneously reflected or
simultaneously transmitted through the barrier. One can show that%
\begin{equation}
S(\hat{\rho}_{N,n}^{(1)})=-\left[ \left\vert R_{n}\right\vert ^{2}\ln
\left\vert R_{n}\right\vert ^{2}+\left\vert T_{n}\right\vert ^{2}\ln
\left\vert T_{n}\right\vert ^{2}\right] \ \text{\textrm{for (a) and (b)}},
\label{mr.4}
\end{equation}%
where the notation in (\ref{RTsqr}) for the reflection and transmission coefficients has been used. Entropy for case 
\textrm{(c)} vanishes, i.e., $S(\hat{\rho}_{N,n}^{(1)})=0$, as $\hat{\rho}_{N,n}^{(1)}$
corresponding to case \textrm{(c)} describes the pure state, despite the fact
that a measurement has been performed in the system. Taking into account
that for this case the differential numbers of final particles are%
\begin{align}
& N_{n,-}(\mathrm{out})=\mathrm{tr}\hat{\rho}_{N,n}^{(1)}\ _{-}a_{n}^{\dag
}\ _{-}a_{n}=\left\vert T_{n}\right\vert ^{2},\ \ \left\vert
R_{n}\right\vert ^{2}=1-\ N_{n,-}(\mathrm{out}),  \notag \\
& N_{n,+}(\mathrm{out})=\mathrm{tr}\hat{\rho}_{N,n}^{(1)}\ ^{+}a_{n}^{\dag
}\ ^{+}a_{n}=\left\vert T_{n}\right\vert ^{2},\ \ \left\vert
R_{n}\right\vert ^{2}=1-\ N_{n,+}(\mathrm{out}),  \label{mr.5}
\end{align}%
we can represent entropies (\ref{mr.4}) as%
\begin{align}
& S(\hat{\rho}_{N,n}^{(1)})=-\left\{ \left[ 1-N_{n,-}(\mathrm{out})\right]
\ln \left[ 1-N_{n,-}(\mathrm{out})\right] +N_{n,-}(\mathrm{out})\ln N_{n,-}(%
\mathrm{out})\right\} \ \ \text{\textrm{for (a),}}  \notag \\
& S(\hat{\rho}_{N,n}^{(1)})=-\left\{ \left[ 1-N_{n,+}(\mathrm{out})\right]
\ln \left[ 1-N_{n,+}(\mathrm{out})\right] +N_{n,+}(\mathrm{out})\ln N_{n,+}(%
\mathrm{out})\right\} \ \ \text{\textrm{for (b).}}  \label{mr.6}
\end{align}%
The entropies for the range $\Omega _{5}$ have the same form,%
\begin{align}
& S(\hat{\rho}_{N,n}^{(5)})=-\left[ \left( 1-N_{n,+}(\mathrm{out})\right)
\ln \left( 1-N_{n,+}(\mathrm{out})\right) +N_{n,+}(\mathrm{out})\ln N_{n,+}(%
\mathrm{out})\right] ,  \notag \\
& S(\hat{\rho}_{N,n}^{(5)})=-\left[ \left( 1-N_{n,-}(\mathrm{out})\right)
\ln \left( 1-N_{n,-}(\mathrm{out})\right) +N_{n,-}(\mathrm{out})\ln N_{n,-}(%
\mathrm{out})\right] ,  \label{mr.7}
\end{align}%
for the cases of single left and single right initial positrons. 

In a similar way and with the same result one can consider the range $\Omega
_{5}$. 

In ranges $\Omega _{2}$, $\Omega _{3}$, and $\Omega _{4}$ all initial
particles are subjected to total reflection; for this reason the
consideration of modes with only one initial particle (electron or
positron) is trivial. The only exception is the case when we have an
initial electron-positron pair in mode $m\in \Omega _{3}$. In this situation
one can show, using Eqs. (\ref{gen.9}), (\ref{gen.14}), and (\ref{V3}) and
relations (\ref{omega.g}) and\ (\ref{gf.dec.prop}), that the initial state
evolves as follows:%
\begin{equation}
\ \ \ ^{-}a_{m}^{\dag }\ _{-}b_{m}^{\dag }\left\vert 0,\text{\textrm{in}}%
\right\rangle _{m}^{(3)}=w_{m}^{\ast }(+|+)^{-1}\left[ \ ^{+}a_{m}^{\dag }\
_{+}b_{m}^{\dag }-w_{m}^{\ast }\left( +-|0\right) \right] \left\vert 0,\text{%
\textrm{out}}\right\rangle _{m}^{(3)}.\   \label{rd.14a}
\end{equation}%
In this expression, the first term on the right-hand side is
the state vector corresponding to the situation when both initial particles
are reflected from the potential step and the second is the vacuum state
vector corresponding to the situation when the initial pair is annihilated. The
partial density operator reduced by measurement of the number of particles with
an initial pair in mode $m\in \Omega _{3}$ has the form 
\begin{equation}
\hat{\rho}_{N,m}^{(3)}\ =\left\vert c_{\mathrm{v},m}\right\vert ^{2}\
\left\vert w_{m}\left( +-|0\right) \right\vert ^{2}\ P_{v,n}^{(3)}(\mathrm{%
out})+\left\vert c_{\mathrm{v},m}\right\vert ^{2}\ \ ^{+}a_{m}^{\dag }\
_{+}b_{m}^{\dag }\ P_{v,n}^{(3)}(\mathrm{out})\ _{+}b_{m}\ ^{+}a_{m},
\label{rd.14c}
\end{equation}%
where $\left\vert c_{\mathrm{v},m}\right\vert ^{2}\ \left\vert w_{m}\left(
+-|0\right) \right\vert ^{2}$ and $\left\vert c_{\mathrm{v},m}\right\vert
^{2}$ are the probabilities of pair annihilation and pair scattering,
respectively. The von Neumann entropy for the density operator $\hat{\rho}%
_{N,m}^{(3)}$ is%
\begin{eqnarray}
&&\ S(\hat{\rho}_{N,m}^{(3)})=-\left[ \left\vert c_{\mathrm{v},m}\right\vert
^{2}\ln \left\vert c_{\mathrm{v},m}\right\vert ^{2}+\left\vert c_{\mathrm{v}%
,m}\right\vert ^{2}\ \left\vert w_{m}\left( +-|0\right) \right\vert ^{2}\ln
\left\vert c_{\mathrm{v},m}\right\vert ^{2}\ \left\vert w_{m}\left(
+-|0\right) \right\vert ^{2}\right]   \notag \\
&&\ =-\left[ \left( 1-N_{n}^{\text{$\mathrm{cr}$}}\right) \ln \left(
1-N_{n}^{\text{$\mathrm{cr}$}}\right) +N_{n}^{\text{$\mathrm{cr}$}}\ln
N_{n}^{\text{$\mathrm{cr}$}}\right] ,  \label{rde.14}
\end{eqnarray}%
the same as for the case of the density operator with the vacuum initial
condition, reduced by the measurement of final particles. 

Now, let us consider a reduction over the subsystems of electrons or
positrons. Due to the nature of this reduction the partial density operators 
$\hat{\rho}_{n}^{(i)}$, $i=1,2,4,5$, either are not affected by the
reduction or are completely traced out. The states with a definite number of
initial particles with fixed quantum numbers $n$ are pure states. This means
that the corresponding entropies vanish, $S(\hat{\rho}_{n}^{(1,2,4,5)})=0$.
In the range $\Omega _{3}$ initial electrons and positrons are subjected to
total reflection \cite{x-case}, i.e., states with a single initial electron
or a single initial positron remain pure states. Then, by applying the procedure
given by Eq. (\ref{nm.1}) to the normal form of the generating functional (%
\ref{gf.7}) it is a simple matter to show that the corresponding entropies $%
S(\hat{\rho}_{\pm ,n}^{(3)})$ vanish. The only interesting result arises
when we consider a mode with an initial electron-positron pair in $\Omega
_{3}$. The partial density operator $\hat{\rho}_{n}^{(3)}$ then takes the
form 
\begin{equation}
\hat{\rho}_{n}^{(3)}=\left\vert \Psi \right\rangle _{n}\ _{n}\left\langle
\Psi \right\vert ,\ \ \left\vert \Psi \right\rangle _{n}=w_{n}^{\ast
}(+|+)^{-1}\left[ \ ^{+}a_{n}^{\dag }\ _{+}b_{n}^{\dag }-w_{n}^{\ast }\left(
+-|0\right) \right] \left\vert 0,\text{\textrm{out}}\right\rangle _{n}^{(3)}
\label{en.7b}
\end{equation}%
and the reduced partial density operators $\hat{\rho}_{\pm ,n}^{(3)}$ can be
calculated as%
\begin{align}
& \hat{\rho}_{+,n}^{(3)}=\left\vert w_{n}(+|+)\right\vert ^{-2}\left[ \
^{+}a_{n}^{\dag }\ P_{+,n}^{(3)}(\text{\textrm{out}})\ ^{+}a_{n}+\left\vert
w_{n}\left( +-|0\right) \right\vert ^{2}\ P_{+,n}^{(3)}(\text{\textrm{out}})%
\right] ,  \notag \\
& \hat{\rho}_{-,n}^{(3)}=\left\vert w_{n}(+|+)\right\vert ^{-2}\left[ \
_{+}b_{n}^{\dag }\ P_{-,n}^{(3)}(\text{\textrm{out}})\ _{+}b_{n}+\left\vert
w_{n}\left( +-|0\right) \right\vert ^{2}\ P_{-,n}^{(3)}(\text{\textrm{out}})%
\right] ,  \label{en.7c}
\end{align}%
where $\left\vert w_{n}(+|+)\right\vert ^{-2}=\left\vert c_{\mathrm{v},n}\right\vert ^{2}$ and projectors $P_{\pm ,n}^{(3)}($\textrm{out}$)$ are 
\begin{equation}
P_{+,n}^{(3)}(\text{\textrm{out}})=\left\vert 0,\text{\textrm{out}}%
\right\rangle _{a,n}^{(3)}\ _{a,n}^{(3)}\left\langle 0,\text{\textrm{out}}%
\right\vert ,\ \ P_{-,n}^{(3)}(\text{\textrm{out}})=\left\vert 0,\text{%
\textrm{out}}\right\rangle _{b,n}^{(3)}\ _{b,n}^{(3)}\left\langle 0,\text{%
\textrm{out}}\right\vert .\   \label{en.7proj}
\end{equation}%
The corresponding von Neumann entropies can be calculated,%
\begin{equation}
S(\hat{\rho}_{\pm ,n}^{(3)})=-\left[ \left\vert c_{\mathrm{v},n}\right\vert
^{2}\ln \left\vert c_{\mathrm{v},n}\right\vert ^{2}+\left\vert c_{\mathrm{v}%
,n}\right\vert ^{2}\left\vert w_{n}\left( +-|0\right) \right\vert ^{2}\ln
\left\vert c_{\mathrm{v},n}\right\vert ^{2}\left\vert w_{n}\left(
+-|0\right) \right\vert ^{2}\right] .  \label{en.reg5}
\end{equation}%
Using the relations (\ref{w.g.n}), we can present it as 
\begin{equation}
S(\hat{\rho}_{\pm ,n}^{(3)})=-\left[ \left( 1-N_{n}^{\text{$\mathrm{cr}$}%
}\right) \ln \left( 1-N_{n}^{\text{$\mathrm{cr}$}}\right) +N_{n}^{\text{$%
\mathrm{cr}$}}\ln N_{n}^{\text{$\mathrm{cr}$}}\right] .  \label{en.reg6}
\end{equation}%
One can see that this result coincides with the one obtained for the case of
initial vacuum state, Eq. (\ref{en.7}), i.e., with the case when an
electron-positron pair is produced from the vacuum, and with the entropy for
the density operator $\hat{\rho}_{N,m}^{(3)}$ given by Eq. (\ref{rde.14}).

\section*{Appendix C}

\appendix\setcounter{equation}{0} \renewcommand{\theequation}{C%
\arabic{equation}} Here we provide several relations that we have used during
the calculations. For both the Fermi and Bose cases the relations  
\cite{density}:%
\begin{align}
& ae^{a^{\dag}Da}=e^{a^{\dag}Da}e^{D}a,\ \ a^{\dag}e^{a^{\dag}Da}=e^{a^{\dag
}Da}a^{\dag}e^{-D},  \label{A.1} \\
& e^{a^{\dag}Da}=:\exp\left\{ a^{\dag}\left( e^{D}-1\right) a\right\} :\ 
\label{A.2}
\end{align}
hold, where in general case $D$ is an arbitrary matrix. Note that in the case
under consideration all products are diagonal in quantum numbers $n$, $%
a^{\dag}Da=a_{n}^{\dag}D_{nn}a_{n}$, and the matrices $D_{nn}$ are diagonal
and single rank, i.e., are just $c$-numbers. One can also easily see that the
following generalization of Eq. (\ref{A.1}) holds:
\begin{equation}
e^{a^{\dag}Da}a=e^{-D}ae^{a^{\dag}Da},\ \
e^{a^{\dag}Da}a^{\dag}=e^{D}a^{\dag }e^{a^{\dag}Da}.  \label{A.3}
\end{equation}
For the product of two normal-form operators the relation
\begin{equation}
:e^{a^{\dag}Da}::e^{a^{\dag}\tilde{D}a}:=:e^{a^{\dag}\left( D+\tilde {D}+D%
\tilde{D}\right) a}:,  \label{A.4}
\end{equation}
is useful, where $D$ and $\tilde{D}$ are matrices, and its simple generalization%
\begin{equation}
:e^{b^{\dag}Da}::e^{a^{\dag}\tilde{D}c}:=:\exp\left[ b^{\dag}Da+a^{\dag }%
\tilde{D}c+b^{\dag}D\tilde{D}c\right] :,  \label{A.5}
\end{equation}
where for the case of Fermi operators the decomposition of the exponent is
finite and has the form%
\begin{equation*}
:\exp\left[ b^{\dag}Da+a^{\dag}\tilde{D}c+b^{\dag}D\tilde{D}c\right]
:=1+b^{\dag}Da+a^{\dag}\tilde{D}c+b^{\dag}D\tilde{D}c-b^{\dag}a^{\dag}Dac.
\end{equation*}
The trace of a normal product of creation and annihilation operators can be
calculated by using the following path integral representation. Suppose that 
$X(a^{\dag },a)$ is an operator expression of creation and annihilation
operators. Then the trace of its normal form can be expressed as \cite%
{density}
\begin{equation}
\text{\textrm{tr}}\left\{ :X(a^{\dag },a):\right\} =\left\langle
0\right\vert \int \exp \left\{ \lambda ^{\ast }\lambda +\lambda ^{\ast
}a\right\} :X(a^{\dag },a):\exp \left\{ a^{\dag }\lambda \right\} \prod
d\lambda ^{\ast }d\lambda \left\vert 0\right\rangle .  \label{A.6}
\end{equation}

\end{document}